\documentclass[prd,showpacs,amsmath,amssymb,twocolumn,floatfix]{revtex4}
\usepackage{graphicx,color,dcolumn}
\begin{document}

\title{Strong decays of charmed baryons}

\author{Chong Chen}
\author{Xiao-Lin Chen}
\author{Xiang Liu}
\email{xiangliu@pku.edu.cn}
\author{Wei-Zhen Deng}
\author{Shi-Lin Zhu}
\email{zhusl@phy.pku.edu.cn}

\affiliation{Department of physics, Peking University, Beijing,
100871, China}

\vspace*{1.0cm}

\date{\today}

\begin{abstract}
There has been important experimental progress in the sector of
heavy baryons in the past several years. We study the strong
decays of the S-wave, P-wave, D-wave and radially excited charmed
baryons using the $^3P_0$ model. After comparing the calculated
decay pattern and total width with the available data, we discuss
the possible internal structure and quantum numbers of those
charmed baryons observed recently.
\end{abstract}

 \maketitle

\section{Introduction}\label{sect0}

Babar and Belle collaborations observed several excited charmed
baryons: $\Lambda_{c}(2880,2940)^+$, $\Xi_c(2980,3077)^{+,0}$ and
$\Omega_{c}(2768)^0$ last year
\cite{babar-2880,belle-2880,babar-2980-3077,belle-2980-3077,babar-omega},
which inspired several investigations of these states in literature
\cite{rosner,xiang,cheng,valcarce}. We collect the experimental
information of these recently observed hadrons in Table \ref{charmed
baryon}. Their quantum numbers have not been determined except
$\Lambda_{c}(2880)^+$. In order to understand their structures using
the present experimental information, we study the strong decay
pattern of the excited charmed baryons systematically in this work.
In the past decades, there has been some research work on heavy
baryons \cite{cheng,charmed baryons,review}.

The quantum numbers and decay widths of S-wave and some P-wave
charmed baryons are known \cite{PDG}. We first systematically
analyze their strong decays in the framework of the $^{3}P_{0}$
strong decay model. Accordingly one can extract the parameters and
estimate the accuracy of the $^{3}P_{0}$ model when it's applied
in the charmed baryon system. Then we go one step further and
extend the same formalism to study the decay patterns of these new
charmed baryons $\Lambda_{c}(2880,2940)^+$, $\Xi(2980,3077)^{+,0}$
under different assignments of their quantum numbers. After
comparing the theoretical results with the available experimental
data, we can learn their favorable quantum numbers and assignments
in the quark model.

\begin{widetext}
\begin{center}
\begin{table}[htb]
\begin{tabular}{c||c|c|ccccccccc} \hline
State&Mass and Width (MeV)&Decay channels in experiments& Other
information\\\hline\hline&$2881.5\pm 0.3$, $<8$
\cite{cleo-2880}&$\Lambda_{c}\pi^{+}\pi^{-}$& \\\cline{2-3}
&$2881.9\pm0.1\pm0.5$ , $5.8\pm1.5\pm1.1$ \cite{babar-2880}&$D^0 p$&
\raisebox{2ex}[0pt]{$J^{P}$ favors $\frac{5}{2}^+$
\cite{belle-2880},}\\\cline{2-3}\raisebox{3ex}[0pt]{$\Lambda_{c}(2880)^{+}$}&$2881.2\pm0.2^{+0.4}_{-0.3}$,
$5.5^{+0.7}_{-0.5}\pm0.4$ \cite{belle-2880}&$\Sigma_{ c}^{\star
0,++}(2520)\pi^{+,-}$&\raisebox{2ex}[0pt]{$\frac{\Gamma(\Sigma_{
c}^{\star}(2520)\pi^{\pm})}
{\Gamma(\Sigma_{c}(2455)\pi^{\pm})}=0.225\pm0.062\pm0.025$
\cite{belle-2880}}
\\\hline\hline
&$2939.\pm1.3\pm1.0$, $17.5\pm5.2\pm5.9$ \cite{babar-2880}&$D^0 p$&\\
\cline{2-3}\raisebox{2ex}[0pt]{$\Lambda_{c}(2940)^+$}&$2937.9\pm1.0^{+1.8}_{-0.4}$,
$10\pm4\pm5$ \cite{belle-2880}&$\Sigma_{c}(2455)^{0,++}\pi^{+,-}$
&\raisebox{2ex}[0pt]{-}\\\hline\hline &$2967.1\pm1.9\pm1.0$,
$23.6\pm2.8\pm1.3$ \cite{babar-2980-3077}&$\Lambda_{c}^{+}K^{-}
\pi^{+}$&\\\cline{2-3}
\raisebox{2ex}[0pt]{$\Xi_{c}(2980)^+$}&$2978.5\pm2.1\pm2.0$,
$43.5\pm7.5\pm7.0$ \cite{belle-2980-3077}&$\Lambda_{c}^{+}K^{-}
\pi^{+}$&\raisebox{2ex}[0pt]{-}\\\hline\hline
$\Xi_{c}(2980)^0$&$2977.1\pm8.8\pm3.5$, $43.5$
\cite{belle-2980-3077}&$\Lambda_{c}^{+}K_{S}^{0}
\pi^{-}$&-\\\hline\hline &$3076.4\pm0.7\pm0.3$, $6.2\pm1.6\pm0.5$
\cite{babar-2980-3077}&$\Lambda_{c}^{+}K^{-}
\pi^{+}$&\\\cline{2-3}
\raisebox{2ex}[0pt]{$\Xi_{c}(3077)^+$}&$3076.7\pm0.9\pm0.5$,
$6.2\pm1.2\pm0.8$ \cite{belle-2980-3077}&$\Lambda_{c}^{+}K^{-}
\pi^{+}$&\raisebox{2ex}[0pt]{-}\\\hline\hline
$\Xi_{c}(3077)^0$&$3082.8\pm1.8\pm1.5$, $5.2\pm3.1\pm1.8$
\cite{belle-2980-3077}&$\Lambda_{c}^{+}K_{S}^{0}
\pi^{-}$&-\\\hline\hline
$\Omega_{c}(2768)^{0}$&$2768.3\pm 3.0$
\cite{babar-omega}&$\Omega_{c}^{0}\gamma$&$J^{P}=\frac{3}{2}^{+}$\\\hline
\end{tabular}

\caption{A summary of recently observed charmed baryons by Babar and
Belle collaborations. \label{charmed baryon}}
\end{table}
\end{center}
\end{widetext}

Very recently CDF collaboration reported four particles
\cite{CDF,CDF-1}, which are consistent with $\Sigma_{b}^{\pm}$ and
$\Sigma_{b}^{*\pm}$ predicted in the quark model \cite{theory}.
Their masses are $ M_{\Sigma_{b}^{+}}=5808^{+2.0}_{-2.3}\pm1.7\;
{\rm MeV}$, $ M_{\Sigma_{b}^{-}}=5816^{+1.0}_{-1.0}\pm1.7 {\rm
MeV}$, $ M_{\Sigma_{b}^{*+}}=5829^{+1.6}_{-1.8}\pm1.7$, $
M_{\Sigma_{b}^{*-}}=5837^{+2.1}_{-1.9}\pm1.7 {\rm MeV}$. The mass
splitting between $\Sigma_{b}$ and $\Sigma_{b}^{*}$ was discussed
in Refs. \cite{rosner-1,lipkin} while the strong decays of
$\Sigma_{b}^{\pm(*)}$ were studied in Ref. \cite{hwang}. As a
byproduct, we also calculate the strong decays of
$\Sigma_{b}^{(*)\pm}$ and other S-wave bottom baryons in this
work.

This paper is organized as follows. We give a short theoretical
review of S-wave, P-wave and D-wave charmed baryons and introduce
our notations for them in Section \ref{sect1}. Then we give a
brief review of $^{3}P_{0}$ model in Section \ref{sect2}. We
present the strong decay amplitudes of charmed baryons in Section
\ref{sect3}. Section \ref{sect4} is the numerical results. The
last section is our discussion and conclusion. Some lengthy
formulae are collected in the Appendix.

\section{The notations and conventions of charmed baryon\label{sect1}}

\begin{figure}[htb]
\begin{center}
\scalebox{0.6}{\includegraphics{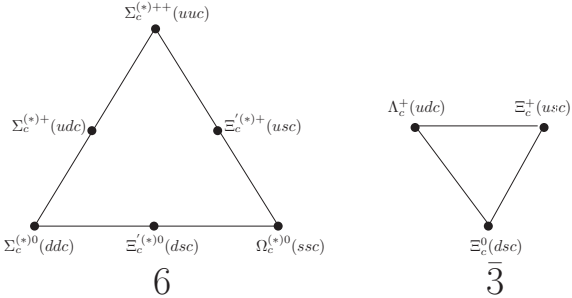}}
\end{center}
\caption{The SU(3) flavor multiplets of charmed
baryons\label{baryon}}
\end{figure}

We first introduce our notations for the excited charmed baryons.
Inside a charmed baryon there are one charm quark and two light
quarks ($u$, $d$ or $s$). It belongs to either the symmetric $6_F$
or antisymmetric $\bar{3}_F$ flavor representation (see Fig.
\ref{baryon}). For the S-wave charmed baryons, the total
color-flavor-spin wave function and color wave function must be
symmetric and antisymmetric respectively. Hence the spin of the
two light quarks is S=1 for $6_F$ or S=0 for $\bar{3}_F$. The
angular momentum and parity of the S-wave charmed baryons are
$J^{P}=\frac{1}{2}^+$ or $\frac{3}{2}^{+}$ for $6_F$ and
$J^{P}=\frac{1}{2}^{+}$ for $\bar{3}_F$. The names of S-wave
charmed baryons are listed in Fig. \ref{baryon}, where we use the
star to denote $\frac{3}{2}^{+}$ baryons and the prime to denote
the $J^P=\frac{1}{2}^{+}$ baryons in the ${6}_F$ representation.

\begin{figure}[htb]
\begin{center}
\scalebox{0.7}{\includegraphics{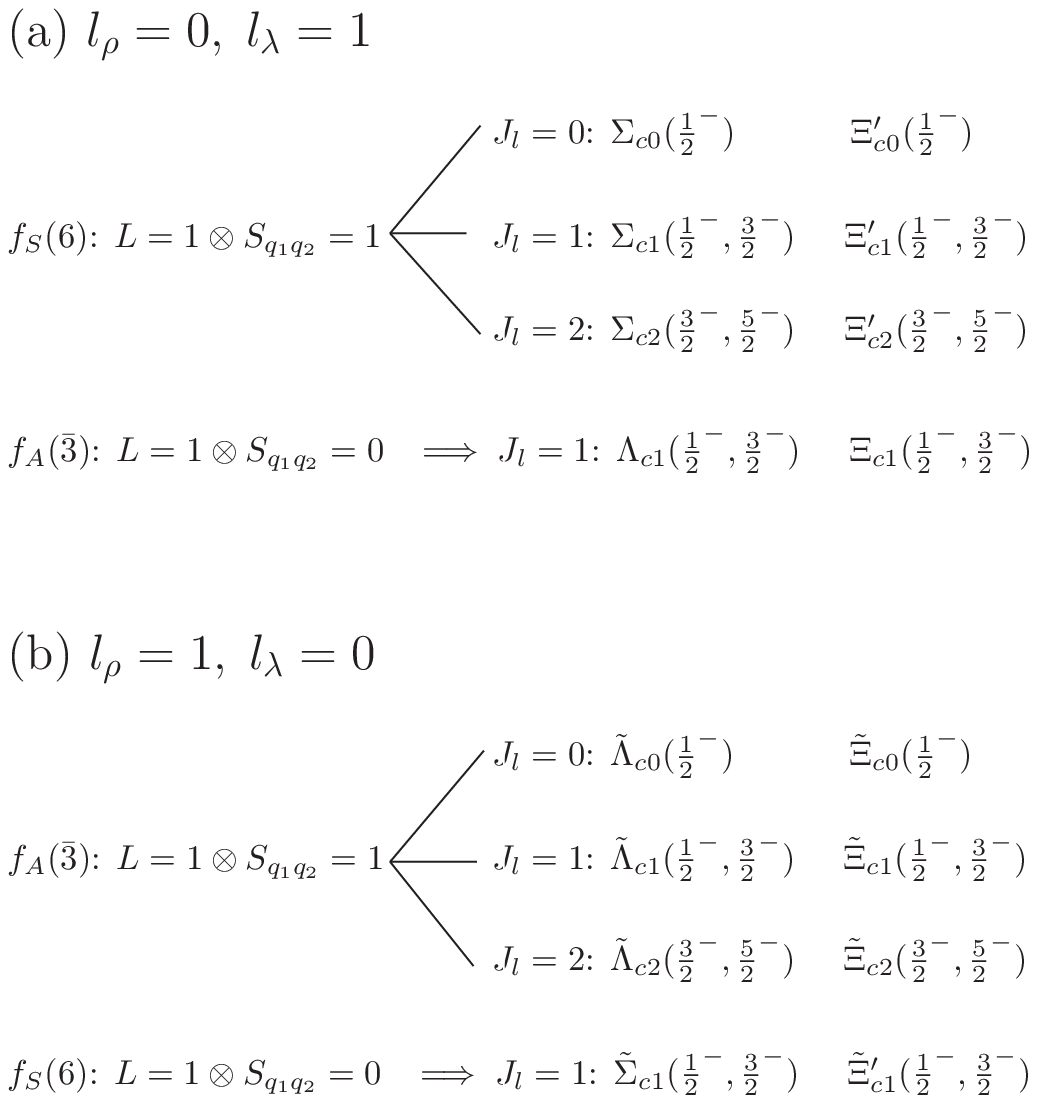}}
\end{center}
\caption{The notations for P-wave charmed baryons. $f_{S}(6_F)$
and $f_{A}({\bar 3}_F)$ denote the SU(3) flavor representation.
$S_{q_{1}q_{2}}$ is the total spin of the two light quarks. $L$
denotes the total orbital angular momentum of charmed baryon
system.\label{p-wave}}
\end{figure}

In Fig. \ref{p-wave} we introduce our notations and conventions
for the P-wave charmed baryons. $l_{\rho}$ is the orbital angular
momentum between the two light quarks while $l_{\lambda}$ denotes
the orbital angular momentum between the charm quark and the two
light quark system. We use the prime to label the $\Xi_{cJ_{l}}$
baryons in the $6_F$ representation and the tilde to discriminate
the baryons with $l_{\rho}=1$ from that with $l_{\lambda}=1$.

\begin{center}
\begin{figure}[htb]
\begin{tabular}{c}
\scalebox{0.7}{\includegraphics{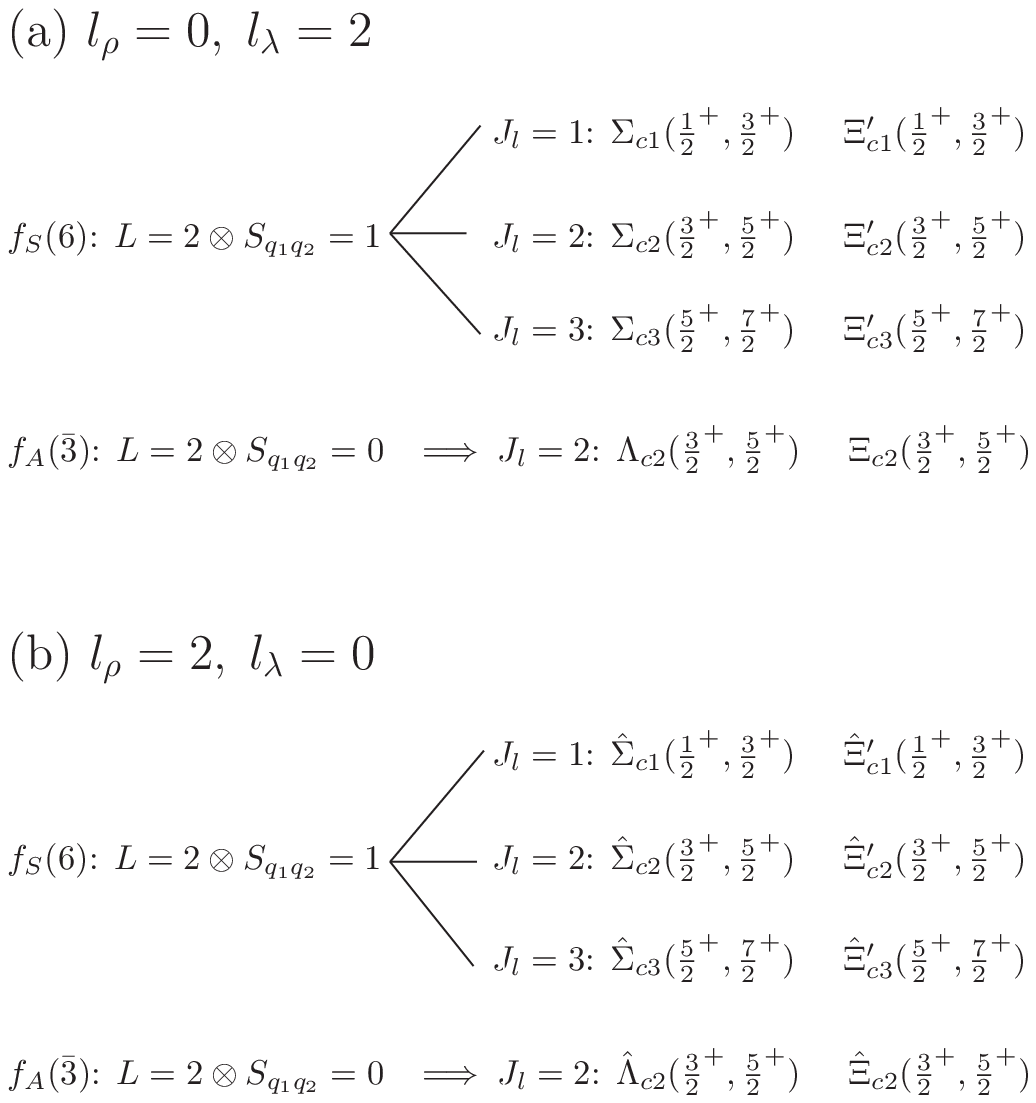}}\\
\\
\scalebox{0.7}{\includegraphics{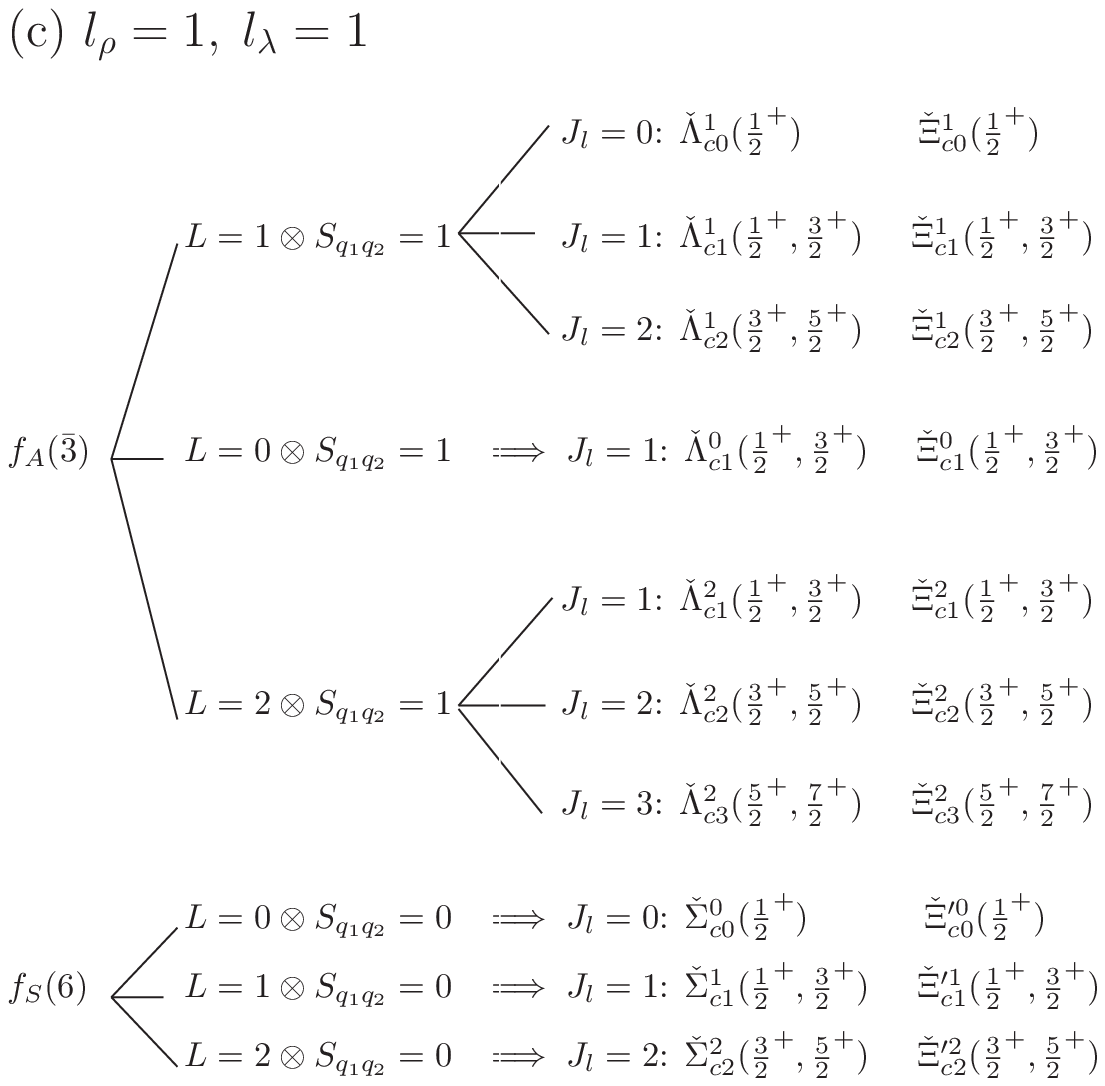}}
\end{tabular}
\caption{The notations for the D-wave charmed
baryons.\label{d-wave}}
\end{figure}\end{center}

The notation for D-wave charmed baryons is more complicated (see
Fig. \ref{d-wave}). Besides the prime, $l_{\rho}$ and
$l_{\lambda}$ defined above, we use the hat and check to denote
the charmed baryons with $l_{\rho}=2$ and $l_{\rho}=1$
respectively. For the baryons with $l_{\rho}=1$ and
$l_{\lambda}=1$, we use the superscript $L$ to denote the
different total angular momentum in
$\check{\Lambda}_{cJ_{l}}^{L}$, $\check{\Sigma}_{cJ_{l}}^{L}$ and
$\check{\Xi}_{cJ_{l}}^{L}$.

\section{The $^3P_{0}$ model\label{sect2}}

The $^{3}P_{0}$ model was first proposed by Micu \cite{Micu} and
further developed by Yaouanc et al. later
\cite{yaouanc,yaouanc-1,yaouanc-book}. Now this model is widely
used to study the strong decays of hadrons
\cite{qpc-1,qpc-2,qpc-90,ackleh,Zou,liu,lujie,baryon-decay}.

\begin{figure}[htb]
\begin{center}
\scalebox{1.5}{\includegraphics{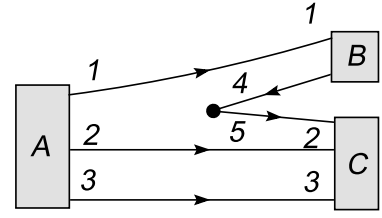}}
\end{center}
\caption{The decay process of $A\to B+C$ in $^3P_{0}$
model.}\label{fig1}
\end{figure}

According to this model, a pair of quarks with $J^{PC}=0^{++}$ is
created from the vacuum when a hadron decays, which is shown in
Fig. \ref{fig1} for the baryon decay process $A\to B+C$. The new
$q\bar{q}$ pair created from the vacuum together with the $qqq$
within the the initial baryon regroup into the outgoing meson and
baryon via the quark rearrangement process. In the
non-relativistic limit, the transition operator is written as
\begin{eqnarray}
T &=& - 3 \gamma \sum_m\: \langle 1\;m;1\;-m|0\;0 \rangle\,
\int\!{\rm d}^3{\textbf{k}}_4\; {\rm d}^3{\textbf{k}}_5
\delta^3({\textbf{k}}_4+{\textbf{k}}_5)\nonumber\\&&\times {\cal
Y}^m_1\Big({{\textbf{k}}_4-{\textbf{k}_5}\over{2}}\Big)\; \chi^{4
5}_{1, -\!m}\; \varphi^{4 5}_0\;\, \omega^{4 5}_0\;
b^\dagger_{4i}({\textbf{k}}_4)\; d^\dagger_{5j}({\textbf{k}}_5)
\label{tmatrix}
\end{eqnarray}
where $i$ and $j$ are the color indices of the created quark and
anti-quark. $\varphi^{45}_{0}=(u\bar u +d\bar d +s \bar s)/\sqrt
3$ and $\omega_{0}^{45}=\delta_{ij}$ for the flavor and color
singlet respectively. $\chi_{{1,-m}}^{45}$ is for the spin triplet
state. $\mathcal{Y}_{1}^{m}(\mathbf{k})\equiv
|\mathbf{k}|Y_{1}^{m}(\theta_{k},\phi_{k})$ is a solid harmonic
polynomial corresponding to the p-wave quark pair. $\gamma$ is a
dimensionless constant related to the strength of the quark pair
creation from the vacuum, which was extracted by fitting to data.
The hadron and meson state are defined as respectively according
to the definition of mock state \cite{mockmeson}
\begin{eqnarray}
&&|A(n_A \mbox{}^{2S_A+1}L_A \,\mbox{}_{J_A M_{J_A}})
({\textbf{P}}_A) \rangle \nonumber\\&=& \sqrt{2 E_A}\:
\!\!\!\!\!\!\sum_{M_{L_A},M_{S_A}}\!\!\! \langle L_A M_{L_A}
S_A M_{S_A} | J_A M_{J_A} \rangle \nonumber\\
&& \times \;\!\!\int\!{\rm d}^3{\textbf{k}}_{1}{\rm
d}^3{\textbf{k}}_{2}{\rm d}^3{\textbf{k}}_{3}
\delta^3({\bf{k_{1}+k_{2}+k_{3}\!-\!P_A}})\nonumber\\&&\times\psi_{n_A
L_A M_{L_A}}\!({\bf{k}}_{1},{\bf{k}}_{2},{\bf{k}}_{3}) \chi^{1 2
3}_{S_A M_{S_A}}\varphi^{1 2 3}_A\omega^{1 2 3}_A
\nonumber\\&&\times|\;q_{1}({\textbf{k}}_{1})
{q}_{2}({\textbf{k}}_{2}){q}_{3}({\textbf{k}}_{3})\rangle,\nonumber\\
\end{eqnarray}
\begin{eqnarray}\label{mockmeson}
&&|B(n_B \mbox{}^{2S_B+1}L_B \,\mbox{}_{J_B M_{J_B}})
({\textbf{P}}_B) \rangle\nonumber\\ &=& \sqrt{2 E_B}\:
 \sum_{M_{L_B},M_{S_B}}\!\!\! \langle L_B M_{L_B}
S_B M_{S_B} | J_B M_{J_B} \rangle \nonumber\\
&& \times \;\!\!\int\!{\rm d}^3{\textbf{k}}_a{\rm
d}^3{\textbf{k}}_b\delta^3({\bf{k_a+k_b\!-\!P_B}})\psi_{n_B L_B
M_{L_B}}\!({\bf{k_a,k_b}})\nonumber\\&&\times\chi^{a b}_{S_B
M_{S_B}}\varphi^{a b}_B\omega^{a b}_B |\;q_a({\textbf{k}}_a)
\bar{q}_b({\textbf{k}}_b)\rangle
\end{eqnarray}
and satisfy the normalization condition
\begin{eqnarray}
\langle A({\textbf{P}}_A)|A({\textbf{P}}'_A) \rangle &=& 2E_A
\delta^3({\textbf{P}}_A-{\textbf{P}}'_A),\nonumber\\ \langle
B({\textbf{P}}_B)|B({\textbf{P}}'_B) \rangle &=& 2E_B
\delta^3({\textbf{P}}_B-{\textbf{P}}'_B)\; .
\end{eqnarray}
The subscripts $1,\;2,\;3$ denote the quarks of parent hadron A.
$a$ and $b$ refer to the quark and antiquark within the meson B
respectively. ${\textbf{k}}_{i}(i=1,2,3)$ are the momentum of
quarks in hadron A. ${\textbf{k}}_a$ and ${\textbf{k}}_b$ are the
momentum of the quark and antiquark in meson B.
${\textbf{P}}_{A(B)}$ represents the momentum of state A(B).
$S_{A(B)}$ and $J_{A(B)}$ denote the total spin and the total
angular momentum of state A(B).

The S-matrix is defined as
\begin{eqnarray}
\langle f|S|i\rangle&=&I-i2\pi \delta(E_f-E_i)\mathcal{M}^{M_{J_A}
M_{J_B} M_{J_C}}\;.
\end{eqnarray}
The helicity amplitude of the process $A\rightarrow B+C$ in the
center of mass frame of meson A is
\begin{eqnarray}
&&{\mathcal{M}}^{M_{J_A} M_{J_B} M_{J_C}}(A\rightarrow BC)
\nonumber\\ &=&\sqrt{8 E_A E_B E_C}\;\;\gamma\!\!\!\!\!\!\!\!\!\!\!
\sum_{\renewcommand{\arraystretch}{.5}\begin{array}[t]{l}
\scriptstyle M_{L_A},M_{S_A},\\
\scriptstyle M_{L_B},M_{S_B},\\
\scriptstyle M_{L_C},M_{S_C},m
\end{array}}\renewcommand{\arraystretch}{1}\!\!\!\!\!\!\!\!
\langle L_A M_{L_A} S_A M_{S_A} | J_A M_{J_A} \rangle
\nonumber\\&&\times \langle L_B M_{L_B} S_B M_{S_B} | J_B M_{J_B}
\rangle \langle L_C M_{L_C} S_C
M_{S_C} | J_C M_{J_C} \rangle \nonumber\\
&& \times  \langle 1\;m;1\;-m|\;0\;0 \rangle\; \langle
\chi^{235}_{S_C M_{S_C}}\chi^{1 4}_{S_B M_{S_B}}  | \chi^{1 23}_{S_A
M_{S_A}} \chi^{45}_{1 -\!m} \rangle \nonumber\\&&\times
\langle\varphi^{235}_C \varphi^{1 4}_B | \varphi^{1 23}_A
\varphi^{45}_0
 \rangle
\;I^{M_{L_A},m}_{M_{L_B},M_{L_C}}({\textbf{p}}) \;
\end{eqnarray}
where the spatial integral
$I^{M_{L_A},m}_{M_{L_B},M_{L_C}}({\textbf{p}})$ is defined as
\begin{eqnarray}
&&I^{M_{L_A},m}_{M_{L_B},M_{L_C}}({\textbf{p}})\nonumber\\&=&\;
\int\!{\rm d}^3{\textbf{k}}_1{\rm d}^3{\textbf{k}}_2{\rm
d}^3{\textbf{k}}_3{\rm d}^3{\textbf{k}}_4 {\rm d}^3{\textbf{k}}_5
\delta^3({\bf{k_4+k_5}})\nonumber\\
&&\times\delta^3({\bf{k_1+k_2+k_{3}-P_{_A}}})\delta^3({\bf{k_1+k_4-P_{_B}}})\nonumber\\
&&\times\delta^3({\bf{k_2+k_3+k_{5}-P_{_C}}})
 \nonumber\\
&&\times\psi^*_{n_B L_B M_{L_B}}\!
({\textbf{k}}_1,{\textbf{k}}_4)\psi^*_{n_C L_C M_{L_C}}\!
({\textbf{k}}_2,{\textbf{k}}_3,{\textbf{k}}_5)\;\nonumber\\&&\times
\psi_{n_A L_A M_{L_A}}\!
({\textbf{k}}_1,{\textbf{k}}_2,{\textbf{k}}_3)\; {\cal
Y}^m_1\big(\frac{{\bf{k_4-k_5}}}{2}\big). \label{integral}
\end{eqnarray}
$\langle  \chi^{235}_{S_C M_{S_C}} \chi^{1 4}_{S_B M_{S_B}}|
\chi^{1 23}_{S_A M_{S_A}} \chi^{45}_{1 -\!m} \rangle$ and $
\langle\varphi^{235}_C \varphi^{1 4}_B | \varphi^{1 23}_A
\varphi^{45}_0  \rangle$ denote the spin and flavor matrix element
respectively.

The decay width of the process $A\to B+C$ is
\begin{eqnarray*}
\Gamma = \pi^2
\frac{{|\textbf{p}|}}{M_A^2}\frac{s}{2J_{A}+1}\sum_{M_{J_A},M_{J_B},M_{J_C}}
\big|\mathcal{M}^{M_{J_A}M_{J_B}M_{J_C}}\big|^2,
\end{eqnarray*}
where $|\textbf{p}|$ is the momentum of the daughter baryon in the
parent's center of mass frame. $s=1/(1+\delta_{BC})$ is a
statistical factor which is needed if $B$ and $C$ are identical
particles.

\section{The strong decays of charmed baryon\label{sect3}}

According to the $^{3}P_{0}$ model, the decay occurs through the
recombination of the five quarks from the initial charmed baryon
and the created quark pair. So there are three ways of regrouping:
\begin{eqnarray}
&{\mathcal{A}}(q_1, q_2, c_3)+\mathcal{P}(q_4,
\bar{q}_5)\rightarrow{\mathcal{B}}(q_2, q_4,
c_3)+{\mathcal{C}}(q_1,
\bar{q}_5),\label{com-1}\\
&{\mathcal{A}}(q_1, q_2, c_3)+\mathcal{P}(q_4,
\bar{q}_5)\rightarrow{\mathcal{B}}(q_1, q_4,
c_3)+{\mathcal{C}}(q_2, \bar{q}_5),\label{com-2}\\
&{\mathcal{A}}(q_1, q_2, c_3)+\mathcal{P}(q_4,
\bar{q}_5)\rightarrow{\mathcal{B}}(q_1, q_2,
q_4)+{\mathcal{C}}(c_3, \bar{q}_5)
\end{eqnarray}
where $q_{i}$ and $c_{3}$ denote the light quark and charm quark
respectively.

When the excited charmed baryon decays into a charmed baryon plus
a light meson as shown in Eq. (\ref{com-1}) and (\ref{com-2}), the
total decay amplitude reads
\begin{widetext}
\begin{eqnarray}
&&M^{M_{J_A}M_{J_B}M_{J_C}}\nonumber\\&&=-2\gamma \sqrt{8E_A E_B
E_C}\sum_{M_{\rho_{A}}}\,\sum_{M_{L_A}}\,\sum_{M_{\rho_B}}\,\sum_{M_{L_B}}\,\sum_{m_1,m_3,m_4,m}\nonumber\\&&\times
\langle \,J_{1 2}\,M_{1 2};\,s_3\,m_3|J_A\,M_{J_A}\rangle \langle
l_{\rho A}m_{\rho A};\,l_{\lambda A}\,m_{\lambda
A}|L_A\,M_{L_A}\rangle \langle L_A M_{L_A};\,S_{1\,2} m_{1\,2}|J_{1
2} M_{1 2}\rangle\nonumber\\&&\times \langle s_1
m_1;\,s_2\,m_2|S_{1\,2} m_{1\,2}\rangle\,\langle J_{1 4}\,M_{1
4};\,s_3\,m_3|J_B\,M_{J_B}\rangle \,\langle l_{\rho B}m_{\rho
B};\,l_{\lambda B}m_{\lambda
B}|L_B\,M_{L_B}\rangle\nonumber\\&&\times \langle L_B
M_{L_B};\,S_{1\,4} m_{1\,4}|J_{1 4} M_{1 4}\rangle \langle s_1
m_1;\,s_4 m_4|S_{1\,4} m_{1\,4}\rangle \langle 1\,m;\, 1\, -m |0
0\rangle\,\langle s_4 m_4; s_5 m_5 |1\,-m \rangle\nonumber\\&&\times
 \langle L_C\, M_{L_C};\,S_C\,M_C|J_C\,M_{J_C} \rangle
 \langle s_2
m_2;\,s_5 m_5|S_C M_C\rangle \times \langle \phi^{1,4,3}_B
\,\phi^{2,5}_C\,|\phi^{4,5}_0\,\phi^{1,2,3}_A\rangle\times
I^{M_{L_A},m}_{M_{L_B},M_{L_C}}({\mathbf{p}}),
\end{eqnarray}
\end{widetext}
where the pre-factor 2 in front of $\gamma$ arises from the fact
that the amplitude from the Eq. (\ref{com-1}) is the same as that
from Eq. (\ref{com-2}).

The overlap integral in the momentum space is
\begin{eqnarray}
&&I^{M_{L_A},m}_{M_{L_B},M_{L_C}}({{\mathbf{p}}})\nonumber\\&&=\delta^3({\mathbf{P}}_B-{\mathbf{P}}_C)
\int {\rm d}^3 {\mathbf{p}}_1 {\rm d}^3 {\mathbf{p}}_2
\psi^{\ast}_B(l_{\rho B},m_{\rho B},l_{\lambda B},m_{\lambda
B})\,\nonumber\\&&\times \psi^{\ast}_C (L_C\,M_{L_C})\,{\cal
Y}^m_1\big(\frac{{\mathbf{p}}_4- {\mathbf{p}}_5}{2}\big)
\psi_A(l_{\rho A},m_{\rho A},l_{\lambda A},m_{\lambda A}).\nonumber\label{spatial}\\
\end{eqnarray}
Since all hadrons in the final states are S-wave in this work, eq.
(\ref{spatial}) can be further expressed as
\begin{eqnarray}
&&I^{M_{L_A},m}_{M_{L_B},M_{L_C}}({{\mathbf{p}}})\nonumber\\
&&=\delta^3({\mathbf{P}}_B-{\mathbf{P}}_C)\Pi(l_{\rho A},m_{\rho
A},l_{\lambda A},m_{\lambda A},m ),
\end{eqnarray}
where we have
used the harmonic oscillator wave functions for both the meson and
baryon. The expressions of $\Pi(l_{\rho A},m_{\rho A},l_{\lambda
A},m_{\lambda A},m )$ for the decays of S-wave, P-wave and D-wave
charmed baryons are collected in the Appendix. We also move the
lengthy expressions of momentum space integration of S-wave,
P-wave and D-wave charmed baryons to the Appendix.

\section{Numerical results\label{sect4}}

The decay widths of charmed baryons from the $^3P_0$ model involve
several parameters: the strength of quark pair creation from
vacuum $\gamma$, the R value in the harmonic oscillator wave
function of meson and the $\alpha_{\rho,\lambda}$ in the baryon
wave functions. We follow the convention of Ref. \cite{Godfrey}
and take $\gamma= 13.4$, which is considered as a universal
parameter in the $^3P_0$ model. The R value of $\pi$ and $K$
mesons is $2.1$ GeV$^{-1}$ \cite{Godfrey} while it's $R=2.3$
GeV$^{-1}$ for the $D$ meson \cite{parameter-2}.
$\alpha_{\rho}=\alpha_{\lambda}=0.5$ GeV for the proton and
$\Lambda$ \cite{baryon-decay}. For S-wave charmed baryons, the
parameters $\alpha_{\rho}$ and $\alpha_{\lambda}$ in the harmonic
oscillator wave functions can be fixed to reproduce the mass
splitting through the contact term in the potential model
\cite{potential}. Their values are $\alpha_{\rho}=0.6$ GeV and
$\alpha_{\lambda}=0.6$ GeV. For P-wave and D-wave charmed baryons,
$\alpha_{\rho}$ and $\alpha_{\lambda}$ are expected to lie in the
range $0.5\sim 0.7$ GeV. In the following, our numerical results
are obtained with the typical values
$\alpha_{\rho}=\alpha_{\lambda}=0.6$ GeV.

The strong decay widths of the S-wave charmed baryons
$\Sigma^{++,+,0}_{c}(2455)$, $\Sigma^{*++,+,0}_{c}(2520)$ and
$\Xi^{*+,0}_c(2645)$ are listed in Table \ref{S-wave}. Accordingly
the decay widths of S-wave bottomed baryons are presented in Table
\ref{S-wave-bottom}. Because $\Xi_{b}$, $\Xi_{b}^{'}$ and
$\Xi_{b}^{*}$ have not been observed so far, their masses are taken
from the theoretical estimate in Ref. \cite{E.Jenkins}, which are
$m_{\Xi_{b}}=5805.7$ MeV, $m_{\Xi_{b}^{'}}=5950$ MeV and
$m_{\Xi_{b}^{*}}=5966.1$ MeV.

The quantum number and internal structure of the following P-wave
charmed baryons $\Lambda^{+}_{c}(2593)$, $\Lambda^{+}_{c}(2625)$,
$\Xi_{c}^{+,0}(2790)$ and $\Xi_{c}^{+,0}(2815)$ are relatively
known experimentally \cite{PDG}. Their strong decay modes and
widths from the $^3P_0$ model are collected in Table \ref{P-wave}.
The quantum number of $\Sigma^{++}_{c}(2800)$ is still unknown
\cite{PDG}. Thus under different P-wave assignments of
$\Sigma^{++}_{c}(2800)$, we present the strong decay widths of its
possible decay modes in Table \ref{2800}. In the heavy quark
limit, the process $\Sigma^{++}_{c}(2800)\to
\Lambda_{c}^{+}\pi^{+}$ is forbidden if $\Sigma^{++}_{c}(2800)$ is
assigned as $\Sigma_{c1}(\frac{1}{2}^{-})$,
$\Sigma_{c1}(\frac{3}{2}^{-})$,
$\tilde{\Sigma}_{c1}(\frac{1}{2}^{-})$ and
$\tilde\Sigma_{c1}(\frac{3}{2}^{-})$, which is observed in our
calculation as can be seen from Table \ref{2800}.

$\Lambda_{c}(2880)^+$ and $\Lambda_{c}(2940)^+$ are observed in
the invariant mass spectrum of $D^{0}p$ \cite{babar-2880}. The
first radial excitation of $\Lambda_c$ does not decay into
$D^{0}p$ from the $^3P_0$ model. Hence the possibility of
$\Lambda_{c}(2880)^+$ and $\Lambda_{c}(2940)^+$ being a radial
excitation is excluded. We calculate their strong decays assuming
they are D-wave charmed baryons. The results are shown in Table
\ref{2880} and \ref{2940}.

With positive parity, $\Xi(2980)^{+,0}$ and $\Xi(3077)^{+,0}$ can
be either the first radially excited charmed baryons or the D-wave
charmed baryons. With different assumptions of their quantum
numbers we present their strong decay widths in Table \ref{2980},
\ref{3077} and Fig. \ref{S-wave-radical-1}.

\begin{center}
\begin{figure}[htb]
\scalebox{0.8}{\includegraphics{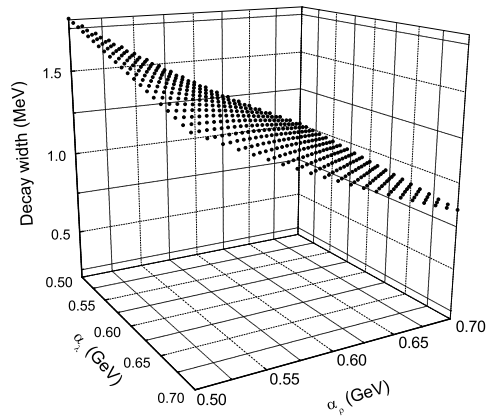}}
\caption{The variation of the decay width of
$\Sigma_{c}^{++}(2455)\to\Lambda_{c}^{+}\pi^{+}$ with
$\alpha_{\rho}$ and $\alpha_{\lambda}$.\label{variation-a}}
\end{figure}\end{center}

\begin{center}
\begin{figure}[htb]
\scalebox{0.8}{\includegraphics{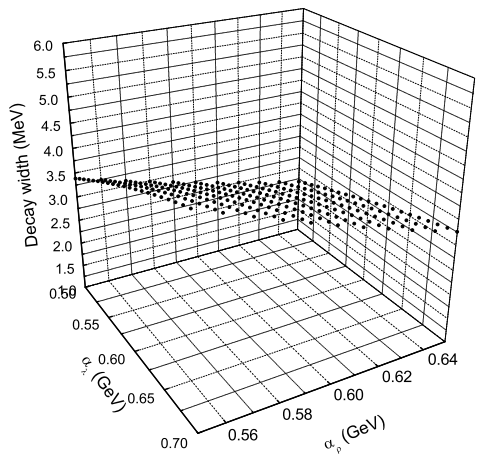}}
\caption{The variation of decay width of
$\Lambda_{c}^{+}(2593)\to\Sigma_{c}^{++}(2455)\pi^{-}$ with
$\alpha_{\rho}$ and $\alpha_{\lambda}$. Here
$\Lambda_{c}^{+}(2593)$ is assigned as
$\Lambda_{c1}(\frac{1}{2}^{-})$. \label{variation-b}}
\end{figure}\end{center}

\begin{center}
\begin{figure}[htb]
\scalebox{0.8}{\includegraphics{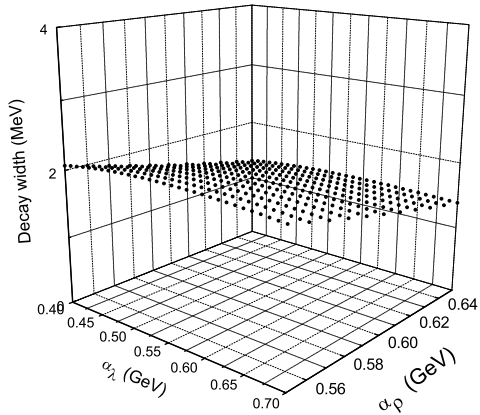}}
\caption{The variation of decay width of
$\Lambda_{c}^{+}(2880)\to\Sigma_{c}^{*++}(2520)\pi^{-}$ with
$\alpha_{\rho}$ and $\alpha_{\lambda}$. Here $\Lambda_{c}^{+}(2880)$
is assigned as $\Lambda_{c2}(\frac{5}{2}^{+})$. \label{variation-c}}
\end{figure}\end{center}

The numerical results depend on the parameters $\alpha_{\rho}$ and
$\alpha_{\lambda}$ in the harmonic oscillator wave functions of
the charmed baryons. We illustrate such a dependence in Figs.
\ref{variation-a}, \ref{variation-b} and \ref{variation-c} using
several typical decay channels:
$\Sigma_{c}^{++}(2455)\to\Lambda_{c}^{+}\pi^{+}$,
$\Lambda_{c}^{+}(2593)\to\Sigma_{c}^{++}(2455)\pi^{-}$ and
$\Lambda_{c}^{+}(2880)\to\Sigma_{c}^{*++}(2520)\pi^{-}$, where
$\Sigma_{c}^{++}(2455)$, $\Lambda_{c}^{+}(2593)$ and
$\Lambda_{c}^{+}(2880)$ are S-wave, P-wave and D-wave baryons
respectively.

\begin{table}[htb]
\begin {center}
\caption{The strong decay widths of S-wave charmed baryons
$\Sigma^{++,+,0}_{c}(2455)$, $\Sigma^{*++,+,0}_{c}(2520)$ and
$\Xi^{*+,0}_c(2645)$. Here all results are in units of MeV.
\label{S-wave}}\vskip 0.3cm
\begin{tabular}{c||ccc|c}
\hline
 &\,\,\,$J^{P}$& \,\,\,Channel  & \,\,Width &\,\, Total width (Exp) \cite{PDG} \\
  \hline\hline
  $\Sigma^{++}_c(2455)$&$\frac{1}{2}^{+}$&$\Lambda^+_c\pi^+ $ &  $1.24$&$2.23\pm0.30$\\
 \hline
$\Sigma^{+}_c(2455)$&$\frac{1}{2}^{+}$&$\Lambda^+_c\pi^0 $ & $1.40$&$<4.6$\\
\hline
$\Sigma^0_c(2455)$&$\frac{1}{2}^{+}$&$\Lambda^+_c\pi^- $ & $1.24$&$2.2\pm0.40$\\
\hline \hline
$\Sigma^{*++}_c(2520)$&$\frac{3}{2}^{+}$&$\Lambda^+_c\pi^+ $ & $11.9$&$14.9\pm1.9$\\
 \hline
$\Sigma^{*+}_c(2520)$&$\frac{3}{2}^{+}$&$\Lambda^+_c\pi^0 $ & $12.1$& $<17$\\
\hline
$\Sigma^{*0}_c(2520)$&$\frac{3}{2}^{+}$&$\Lambda^+_c\pi^- $ & $11.9$&$16.1\pm2.1$\\
\hline \hline
$\Xi^{*+}_c(2645)$&$\frac{3}{2}^{+}$&$\Xi^+_c\pi^0$&$0.64$&\\
\cline{1-4}
$\Xi^{*+}_c(2645)$&$\frac{3}{2}^{+}$&$\Xi^0_c\pi^+$&$0.49$&\raisebox{2ex}[0pt]{$<3.1$}\\
\hline\hline
$\Xi^{*0}_c(2645)$&$\frac{3}{2}^{+}$&$\Xi^+_c\pi^-$&$0.54$&\\
 \cline{1-4}
 $\Xi^{*0}_c(2645)$&$\frac{3}{2}^{+}$&$\Xi^0_c\pi^0$&$0.54$&\raisebox{2ex}[0pt]{$<5.5$}\\
\hline\hline
\end{tabular}
\end{center}
\end{table}
\begin{table}[htb]
\begin {center}
\caption{The strong decay widths of S-wave bottom baryons
$\Sigma_{b}$, $\Sigma^{*}_{b}$, $\Xi'_b$ and $\Xi^{*}_b$. Here all
results are in units of MeV. \label{S-wave-bottom}}\vskip 0.3cm
\begin{tabular}{c||ccc|c}
\hline
 &\,\,\,$J^{P}$& \,\,\,Channel  & \,\,Width &\,\, Experimental results \cite{CDF} \\
  \hline\hline
$\Sigma^{+}_b$&$\frac{1}{2}^{+}$&$\Lambda^0_b\pi^+ $ & $3.5 $&$ $\\
\cline{1-4}
$\Sigma^-_b$&$\frac{1}{2}^{+}$&$\Lambda^0_b\pi^- $ & $ 4.7$&\raisebox{2ex}[0pt]{$\sim 8$}\\
\hline \hline
$\Sigma^{*+}_b$&$\frac{3}{2}^{+}$&$\Lambda^0_b\pi^+ $ & $ 7.5$&$ $\\
 \cline{1-4}
$\Sigma^{*-}_b$&$\frac{3}{2}^{+}$&$\Lambda^0_b\pi^- $ & $ 9.2$&\raisebox{2ex}[0pt]{$\sim 15$}\\
\hline \hline
$\Xi'_b$&$\frac{1}{2}^{+}$&$\Xi_b\pi$&$0.10 $&-\\
\cline{1-5}
$\Xi^{*}_b$&$\frac{3}{2}^{+}$&$\Xi_b\pi$&$0.85$&-\\
\hline\hline
\end{tabular}
\end{center}
\end{table}
\begin {center}
\begin{table}[htb]
\caption{The decay widths of P-wave charmed baryons
$\Lambda^{+}_{c}(2593,2625)$ and $\Xi_{c}^{+,0}(2790,2815)$ with
the fixed structure and quantum number assignments. Here all
results are in units of MeV. \label{P-wave}}\vskip 0.3cm
\begin{tabular}{c||c|cc|c}
  \hline
&Assignment&  Channel & $\Gamma$ & $\Gamma_{Exp}$  \cite{PDG}\\
  \hline
  &&$\Sigma^{++}_{c}\pi^- $ & $3.4$&$$\\
 \cline{3-4}
&&$\Sigma^{+}_{c}\pi^0 $ & $6.4$&$$\\
\cline{3-4}
\raisebox{3ex}[0pt]{$\Lambda^+_c(2593)$}&\raisebox{3ex}[0pt]
{$\Lambda_{c1}(\frac{1}{2}^{-})$}&$\Sigma^0_c\pi^+ $ & $3.4$&\raisebox{2ex}[0pt]{$3.6^{+2.0}_{-1.3}$}\\
\hline \hline
&&$\Sigma^{++}_c\pi^- $ & $1.9\times10^{-3}$&$<0.10$\\
 \cline{3-5}
&&$\Sigma^{+}_c\pi^0 $ & $2.6\times10^{-3}$&$<1.9$\\
\cline{3-5}
\raisebox{3ex}[0pt]{$\Lambda^+_c(2625)$}&\raisebox{3ex}[0pt]{$\Lambda_{c1}(\frac{3}{2}^{-})$}&
$\Sigma^0_c\pi^+ $ & $1.9\times10^{-3}$&$<0.10$\\
\hline \hline
&&$\Xi'^+_c\pi^0$&$5.0$&\\
\cline{3-4}
\raisebox{1.5ex}[0pt]{$\Xi^+_c(2790)$}&\raisebox{1.5ex}[0pt]{$\Xi_{c1}(\frac{1}{2}^{-})$}
&$\Xi'^0_c\pi^+$&$4.9$&\raisebox{1.5ex}[0pt]{$<15$}\\
\hline
&&$\Xi'^+_c\pi^-$&$5.2$&\\
 \cline{3-4}
\raisebox{1.5ex}[0pt]{$\Xi^0_c(2790)$}&\raisebox{1.5ex}[0pt]{$\Xi_{c1}(\frac{1}{2}^{-})$}
&$\Xi'^0_c\pi^0$&$5.1$&\raisebox{1.5ex}[0pt]{$<12$}\\
 \hline
 \hline
&&$\Xi^{\star+}_c\pi^0$&$2.7$&\\
\cline{3-4} \raisebox{1.5ex}[0pt]{$\Xi^+_c(2815)$}
&\raisebox{1.5ex}[0pt]{$\Xi_{c1}(\frac{3}{2}^-)$}&$\Xi^{\star0}_c\pi^+$&$2.6$
&\raisebox{1.5ex}[0pt]{$<3.5$}\\
\hline
&&$\Xi^{\star+}_c\pi^-$&$2.7$&\\
 \cline{3-4}
\raisebox{1.5ex}[0pt]{$\Xi^0_c(2815)$}&\raisebox{1.5ex}[0pt]
{$\Xi_{c1}(\frac{3}{2}^-)$}&$\Xi^{\star0}_c\pi^0$&$2.8$&\raisebox{1.5ex}[0pt]{$<6.5$}\\
 \hline
\hline
\end{tabular}
\end{table}
\end{center}

\begin{table}[htb] 
\caption{The decay widths of $\Sigma^{++}_{c}(2800)$ in different
P-wave charmed baryons assignments. $\mathcal{R}={\Sigma^{+,++}_{c}
\pi^{+,0}}/{\Sigma^{\star+,++}_{c}\pi^{+,0}}$. The total width of
$\Sigma^{++}_{c}(2800)$ is $75^{+22}_{-17}$ MeV \cite{PDG}. Here all
results are in units of MeV. \label{2800}}\vskip 0.3cm
\begin{tabular}{c||cccccccc}\hline
Assignment & \,$\Lambda^{+}_{c} \pi^{+}$&\, $\Sigma^{+,++}_{c}
\pi^{+,0}$&\,
$\Sigma^{\star+,++}_{c}\pi^{+,0}$&\,$\mathcal{R}$\\\hline\hline

$\Sigma_{c0}(\frac{1}{2}^{-})$&$307$&$0.0$&$ 0.0 $&- \\

$\Sigma_{c1}(\frac{1}{2}^{-})$&$ 0.0 $&$296 $&$ 0.4 $&740\\

$\Sigma_{c1}(\frac{3}{2}^{-})$ &$ 0.0 $&$ 0.7 $&$ 220 $&$3\times10^{-3}$\\

$\Sigma_{c2}(\frac{3}{2}^{-})$&$ 8.1 $&$ 1.3  $&$ 0.3 $ &4.3\\

$\Sigma_{c2}(\frac{5}{2}^{-})$ &$ 8.1 $&$ 0.6 $&$ 0.5 $&1.2\\
 \hline
$\tilde{\Sigma}_{c1}(\frac{1}{2}^{-})$&$0.0 $&$  75 $&$ 69 $&1.1\\
$\tilde{\Sigma}_{c1}(\frac{3}{2}^{-})$&$ 0.0  $&$ 75 $&$ 69  $&1.1\\
\hline \hline
\end{tabular}
\end{table}

\section{Discussion and conclusion\label{sect5}}

At present it is still too difficult to calculate the strong decay
widths of hadrons from the first principles of QCD. For this
purpose, some phenomenological strong decay models were proposed
such as the $^{3}P_{0}$ model, flux tube model, QCD sum rule,
lattice QCD etc, among which only the first two approaches can be
applied to the strong decays of excited hadrons.  To a large
extent, the predictions from the $^{3}P_{0}$ and flux tube models
roughly agree with each other.

The $^{3}P_{0}$ model possesses inherent uncertainties
\cite{yaouanc,Godfrey,liu}. In certain cases, the result from the
$^{3}P_{0}$ model may be a factor of $2\sim 3$ off the
experimental width. The uncertainty source of the $^{3}P_{0}$
model arises from the strength of the quark pair creation from the
vacuum $\gamma$, the approximation of non-relativity, and assuming
the simple harmonic oscillator radial wave functions for the
hadrons. Even with the above uncertainty, the $^{3}P_{0}$ model is
still the most systematic, effective and widely used framework to
study the hadron strong decays.

In this work, we have calculated the strong decay widths of
charmed baryons using the $^{3}P_{0}$ model. Our numerical results
do not strongly depend on the parameters $\alpha_{\rho}$ and
$\alpha_{\lambda}$ as shown in Figs. \ref{variation-a},
\ref{variation-b} and \ref{variation-c}. Thus the following
qualitative features and conclusions remain essentially unchanged
with reasonable variations of $\alpha_{\rho}$ and
$\alpha_{\lambda}$.

Our results for the S-wave charmed baryons
$\Sigma^{++,+,0}_{c}(2455)$, $\Sigma^{*++,+,0}_{c}(2520)$ and
$\Xi^{*+,0}_c(2645)$ are roughly consistent with experimental data
within the inherent uncertainty of the $^{3}P_{0}$ model. As a
byproduct, we have also calculated the strong decays of
$\Sigma_{b}^{\pm}$ and $\Sigma_{b}^{*\pm}$ observed by CDF
Collaboration recently. The numerical results are consistent with
the experimental values too.

The decay width of P-wave baryon $\Lambda^{+}_{c}(2593)$ is three
times larger than the experimental value. With the large
experimental uncertainty and the inherent theoretical uncertainty
of the the $^{3}P_{0}$ model, such a deviation is still
acceptable. The decay widths of $\Lambda^{+}_{c}(2625)$ and
$\Xi_{c}^{+,0}(2790,2815)$ are compatible with the experimental
upper bound. By comparing our results with the experimental total
width, we tend to exclude the $\Sigma_{c0}(\frac{1}{2}^{-})$
assignment for $\Sigma_{c}^{++}(2800)$. Since the
$\Sigma_{c}^{++}(2800)$ is observed in $\Lambda_{c}^{+}\pi^{+}$
channel \cite{belle-2800}, there are only two assignments left for
$\Sigma_{c}^{++}(2800)$, i.e. $\Sigma_{c2}(\frac{3}{2}^{-})$ or
$\Sigma_{c2}(\frac{5}{2}^{-})$. More experimental information such
as the ratio ${\Gamma [\Sigma_{c}^{++}(2800)\to \Sigma^{+,++}_{c}
\pi^{+,0}]\over \Gamma [\Sigma_{c}^{++}(2800)\to
\Sigma^{\star+,++}_{c}\pi^{+,0}]}$ will be helpful in the
determination of the quantum number of $\Sigma_{c}^{++}(2800)$.

We have also calculated the strong decay widths of newly observed
$\Lambda_{c}(2880,2940)^+$, $\Xi(2980,3077)^{+,0}$ assuming they
are candidates of D-wave charmed baryons. We find that the only
possible assignment of $\Lambda_{c}(2880)^{+}$ is
$\check{\Lambda}^{2}_{c3}(\frac{5}{2}^{+})$ after considering both
its total decay width and the ratio ${\Gamma(\Sigma^{\star}_{c}
\pi^{\pm})}/{\Gamma(\Sigma_{c} \pi^{\pm})}$, which agrees very
well with the indication from Belle experiment that
$\Lambda_{c}(2880)^{+}$ favors $J^{P}=\frac{5}{2}^{+}$ by the
analysis of the angular distribution \cite{belle-2880}.

Unfortunately the experiment information about the
$\Lambda_{c}(2940)^+$, $\Xi(2980,3077)^{+,0}$ is scarce at
present. From their calculated decay widths, we can only exclude
some assignments which are marked with crosses in Tables
\ref{2940}, \ref{2980} and \ref{3077}. The decay width ratios of
$\Lambda_{c}(2940)^+$, $\Xi(2980,3077)^{+,0}$ from the $^{3}P_{0}$
model will be useful in the identification of their quantum
numbers in the future since the inherent uncertainty cancels
largely.

We have also discussed the strong decays of $\Xi(2980,3077)^{+,0}$
assuming they are radial excitations. Unfortunately the numerical
results in Fig. \ref{S-wave-radical-1} depend quite strongly on
the node of the spatial wave function which is related to the
parameters of the harmonic oscillator wave functions as shown in
Fig. \ref{S-wave-radical-1}. We are unable to make strong
conclusions here.

\section*{Appendix}

\subsection{The harmonic oscillator wave functions used in our
calculation}

For the S-wave charmed baryon,
\begin{eqnarray}
\psi(0,0,0,0)=3^{3/4}\;(\frac{1}{\pi
\alpha^2_{\rho}})^{\frac{3}{4}}\,(\frac{1}{\pi
\alpha^2_{\lambda}})^{\frac{3}{4}}\,\exp\Big[{-\frac{
{\mathbf{p}}^2_{\rho}}{2\alpha^2_{\rho}} -\frac{
{\mathbf{p}}^2_{\lambda}}{2\alpha^2_{\lambda}}}\Big].\nonumber\\
\end{eqnarray}

For the P-wave charmed baryon,
\begin{eqnarray}
\psi(1,m,0,0)&=&-i\;3^{3/4}\,\Big(\frac{8}{3\sqrt{\pi}}\Big)^{{1}/{2}}\Big({1}/{
\alpha^2_{\rho}}\Big)^{ {5}/{4}}\,{\cal
Y}^m_1({\mathbf{p}}_{\rho})\nonumber\\&&\times\Big(\frac{1}{\pi
\alpha^2_{\lambda}}\Big)^{{3}/{4}}\,\exp\Big[{-\frac{{\mathbf{p}}^2_{\rho}}{2\alpha^2_{\rho}}
-\frac{{\mathbf{p}}^2_{\lambda}}{2\alpha^2_{\lambda}}}\Big],\\
\psi(0,0,1,m)&=&-i\;3^{3/4}\,\Big(\frac{8}{3\sqrt{\pi}}\Big)^{{1}/{2}}\Big(\frac{1}{
\alpha^2_{\lambda}}\Big)^{{5}/{4}}\,{\cal
Y}^m_1({\mathbf{p}}_{\lambda})\nonumber\\&&\times \Big(\frac{1}{\pi
\alpha^2_{\rho}}\Big)^{{3}/{4}}\,\exp\Big[{-\frac{{\mathbf{p}}^2_{\rho}}{2\alpha^2_{\rho}}
-\frac{{\mathbf{p}}^2_{\lambda}}{2\alpha^2_{\lambda}}}\Big].
\end{eqnarray}

For the D-wave charmed baryon,
\begin{eqnarray}
\psi
(2,m,0,0)&=&3^{3/4}\;\Big(\frac{16}{15\sqrt{\pi}}\Big)^{{1}/{2}}\Big(\frac{1}{
\alpha^2_{\rho}}\Big)^{{7}/{4}}\,{\cal
Y}^m_2({\mathbf{p}}_{\rho})\nonumber\\&&\times\Big(\frac{1}{\pi
\alpha^2_{\lambda}}\Big)^{{3}/{4}}\,\exp\Big[{-\frac{{\mathbf{p}}^2_{\rho}}{2\alpha^2_{\rho}}
-\frac{{\mathbf{p}}^2_{\lambda}}{2\alpha^2_{\lambda}}}\Big],\\
\psi(0,0,2,m)&=&3^{3/4}\;\Big(\frac{16}{15\sqrt{\pi}}\Big)^{{1}/{2}}\Big(\frac{1}{
\alpha^2_{\lambda}}\Big)^{{7}/{4}}\,{\cal
Y}^m_2({\mathbf{p}}_{\lambda})\nonumber\\&&\times\Big(\frac{1}{\pi
\alpha^2_{\rho}}\Big)^{{3}/{4}}\,\exp\Big[{-\frac{{\mathbf{p}}^2_{\rho}}{2\alpha^2_{\rho}}
-\frac{ {\mathbf{p}}^2_{\lambda}}{2\alpha^2_{\lambda}}}\Big],\\
\psi(1,m,1,m')&=&-3^{3/4}\;\Big(\frac{8}{3\sqrt{\pi}}\Big)^{{1}/{2}}\Big(\frac{1}{
\alpha^2_{\rho}}\Big)^{{5}/{4}}\,{\cal Y}^m_1(
{\mathbf{p}}_{\rho})\nonumber\\&&\times\Big(\frac{8}{3\sqrt{\pi}}\Big)^{{1}/{2}}\Big(\frac{1}{
\alpha^2_{\lambda}}\Big)^{{5}/{4}}{\cal
Y}^{m'}_1({\mathbf{p}}_{\lambda})\nonumber\\&&\times
\exp\Big[{-\frac{{\mathbf{p}}^2_{\rho}}{2\alpha^2_{\rho}}
-\frac{{\mathbf{p}}^2_{\lambda}}{2\alpha^2_{\lambda}}}\Big].
\end{eqnarray}
Here $\mathcal{Y}_{l}^{m}(\mathbf{p})$ is the solid harmonic
polynomial.

The ground state wave function of meson is
\begin{eqnarray}
\psi(0,0)=\Big(\frac{{R}^2}{\pi}\Big)^{{3}/{4}} \exp
\Big[-\frac{R^2({\mathbf{p}}_2-{\mathbf{p}}_5)^2}{8}\Big].
\end{eqnarray}

The the wave function of the first radially excited charmed baryon
$\psi(n_\rho, n_\lambda)$ reads as
\begin{eqnarray*}
&&\psi(1,0)\nonumber\\&&=3^{3/4}\sqrt{\frac{2}{3}}\;\Big(\frac{1}{\pi^{2}
\alpha_{\rho}\alpha_{\lambda}}\Big)^{\frac{3}{2}}
\Big[\frac{3}{2}-\frac{{\mathbf{p}}^2_{\rho}}{\alpha^{2}_{\rho}}\Big]\exp\Big[{-\frac{
{\mathbf{p}}^2_{\rho}}{2\alpha^2_{\rho}} -\frac{
{\mathbf{p}}^2_{\lambda}}{2\alpha^2_{\lambda}}}\Big],\\
&&\psi(0,1)\nonumber\\&&=3^{3/4}\sqrt{\frac{2}{3}}\;\Big(\frac{1}{\pi^{2}
\alpha_{\rho}\alpha_{\lambda}}\Big)^{\frac{3}{2}}
\Big[\frac{3}{2}-\frac{{\mathbf{p}}^2_{\lambda}}{\alpha^{2}_{\lambda}}\Big]\exp\Big[{-\frac{
{\mathbf{p}}^2_{\rho}}{2\alpha^2_{\lambda}} -\frac{
{\mathbf{p}}^2_{\lambda}}{2\alpha^2_{\lambda}}}\Big],
\end{eqnarray*}
where $n_{\rho}$ and $n_{\lambda}$ denote the radial quantum number
between the two light quarks and between heavy quark and the two
light quarks respectively. Here
${\mathbf{p}}_{\rho}=\sqrt{\frac{1}{2}}({\mathbf{p}}_1-{\mathbf{p}}_2)$
and
$\mathbf{p}_{\lambda}=\sqrt{\frac{1}{6}}(\mathbf{p}_1+\mathbf{p}_2-2\mathbf{p}_3)$
for the above expressions. All the above harmonic oscillator wave
functions can be normalized as $\int d\mathbf{p}_{1}d
\mathbf{p}_{2}d \mathbf{p}_{3}|\psi|^2=1$.

\subsection{The momentum space integration}

The momentum space integration $\Pi(l_{\rho A},m_{\rho
A},l_{\lambda A},m_{\lambda A},m)$ includes:

For the S-wave charmed baryon decay,
\begin{eqnarray}
 \Pi(0,0,0,0,0)=\beta |\textbf{p}|\;\Delta_{0,0}.
 \end{eqnarray}

For the P-wave charmed baryon decay,
\begin{eqnarray*}
\Pi(0,0,1,0,0)&=&\frac{1}{2f_1}\big[
f_2 \beta |\textbf{p}|^2-\zeta\big]\;\Delta_{0,1},\\
\Pi(0,0,1,1,-1)&=&\Pi(0,0,1,-1,1)=\frac{\zeta}{2f_1} \;\Delta_{0,1},\\
\nonumber\\
\Pi(1,0,0,0,0)&=&\Big[{\beta \varpi
|\textbf{p}|^2}+\frac{1}{2\sqrt{2}\lambda_1}+\frac{\lambda_2
\zeta}{4\lambda_1 f_1}\Big]\;\Delta_{1,0},\nonumber\\
\Pi(1,1,0,0,-1)&=&\Pi(1,-1,0,0,1)={\beta \varpi
|\textbf{p}|^2}\;\Delta_{1,0}.
\end{eqnarray*}

For the D-wave charmed baryon decay,
\begin{eqnarray*}
 \Pi(0,0,2,0,0)&=&-\frac{f_2}{f_1^2}\Big[\frac{
f_2}{2}\beta \,|\textbf{p}|^3+\zeta |\textbf{p}|\Big]\times\Delta_{0,2},\\
\Pi(0,0,2,1,-1)&=&\Pi(0,0,2,-1,1)=\frac{\sqrt{3}\zeta f_2}{2 f_1^2}
|\textbf{p}|
\Delta_{0,2},\\
\Pi(2,0,0,0,0)&=&-2\Big[(\beta \varpi^{2}|\textbf{p}|^{3}+\frac{1
}{\sqrt{2}\lambda_1}\varpi\,|\textbf{p}|\nonumber\\&&+\frac{\lambda_2
}{2\lambda_1 f_1}\zeta\,\varpi)\Big]\Delta_{2,0},
\\\Pi(2,1,0,0,-1)&=&\Pi(2,-1,0,0,1)\nonumber\\&=&\Big[(\frac{1}{\sqrt{2}\lambda_1}+\frac{\lambda_2
}{2\lambda_1 f_1}\zeta)\Big]\Delta_{2,0}, \\
\Pi(1,0,1,0,0)&=&\Big[\frac{f_2}{2 f_1}\beta \varpi
|\textbf{p}|^3+\frac{1}{2 f_1}(\frac{\lambda_2 \zeta }{2\lambda_1}
\beta+ \zeta \varpi\nonumber\\&&+\frac{\lambda_2 f_2}{4\lambda_1
f_1})|\textbf{p}|+\frac{f_2
}{4\sqrt{2}\lambda_1 f_1}|\textbf{p}|\Big]\Delta_{1,1},\\
\Pi(1,1,1,-1,0)&=&\Pi(1,-1,1,1,0)=\Big[\frac{
\lambda_2}{4\lambda_1 f_1}\beta |\textbf{p}|\Big]\Delta_{1,1},\\
\Pi(1,0,1,1,-1)&=&\Pi(1,0,1,-1,1)=\Big[\frac{1}{2 f_1}\varpi\,\zeta
|\textbf{p}|\Big]\Delta_{1,1},\\
\Pi(1,1,1,0,-1)&=&\Pi(1,-1,1,0,1)\nonumber\\&=&\Big[(\frac{1}{2
\sqrt{2}\lambda_1} +\frac{\lambda_2}{4\lambda_1
f_1}\zeta)\times\frac{f_2}{2 f_1}|\textbf{p}|\Big]\Delta_{1,1}.
\end{eqnarray*}

For the strong decay of the radial excitation, the momentum space
integrals denoted as $\Pi(n_{\rho}, n_{\lambda})$ are:
\begin{eqnarray*}
\Pi(0, 1)&=&\sqrt{\frac{2}{3}}\Big[-\frac{\beta
f^2_2}{4\alpha^2_{\lambda}f^2_1}k^3+\frac{3\beta
}{2}k-\frac{3\beta}{2f_1\alpha^2_{\lambda}}k\\&&+\frac{f_2\zeta}{2f^2_1\alpha^2_{\lambda}}k\Big]\Delta_{0,0},\\
\Pi(1,0)&=&\sqrt{\frac{2}{3}}\Big[\frac{1}{\alpha^2_{\rho}}
(\beta\varpi^2k^3-\frac{3\beta\alpha^2_{\rho}}{2}k+\frac{3\beta}{2\lambda_1}k
\\&&-\frac{\sqrt{2}\varpi}{2\lambda_1}k-\frac{\lambda_2\varpi\zeta}{3\lambda_1}k+\frac{\lambda^2_2}{4\lambda^2_1})\Big]\Delta_{0,0}.
\end{eqnarray*}
where
\begin{eqnarray*}
\lambda_1&=&\frac{1}{\alpha^2_{\rho}}+\frac{1}{4}R^2,\;\;\;
\lambda_2=-\frac{1}{2\sqrt{3}}R^2,\\
\lambda_3&=&\frac{1}{\alpha^2_{\lambda}}+\frac{1}{12}R^2,\;\;\;
\lambda_4=\frac{1}{\sqrt{2}\alpha^2_{\rho}}+\frac{1}{2\sqrt{2}}R^2,\\
\lambda_5&=&-(\frac{1}{\sqrt{6}\alpha^2_{\lambda}}+\frac{1}{2\sqrt{6}}R^2),\\
\lambda_6&=&\frac{1}{4\alpha^2_{\rho}}+\frac{1}{12\alpha^2_{\lambda}}+\frac{1}{8}R^2,\\
f_1&=&\lambda_3-\frac{\lambda_2^2}{4\lambda_1},\;\;\;f_2=\lambda_5-\frac{2\lambda_2\lambda_4}{4\lambda_1},\\
f_3&=&\lambda_6-\frac{\lambda_4^2}{4\lambda_1},\;\;\;
\zeta=\frac{\lambda_2}{2\sqrt{2}\lambda_1}+\frac{1}{\sqrt{6}},\\
\varpi&=&\frac{\lambda_2 f_2}{4\lambda_1 f_1}
-\frac{\lambda_4}{2\lambda_1},\\\beta&=&(1+\frac{\sqrt{3}\lambda_2f_2-2\sqrt{3}
\lambda_4f_1+2\lambda_1f_2}{4\sqrt{6}\lambda_1f_1}),\\
\end{eqnarray*}
and
\begin{eqnarray*}
\Delta_{0,0}&=&(\frac{1}{\pi
\alpha^2_{\rho}})^{\frac{3}{4}}\,(\frac{1}{\pi
\alpha^2_{\lambda}})^{\frac{3}{4}}(\frac{R^2}{\pi})^{\frac{3}{4}}(\frac{\pi^2}{\lambda_1f_1})^{\frac{3}{2}}
\nonumber\\&&\times\exp\Big[-(f_3-\frac{f^2_2}{4f_1})|\textbf{p}|^2\Big]\nonumber\\&&\times\Big[-\sqrt{\frac{3}{4\pi}}(\frac{1}{\pi
\alpha^2_{\rho}})^{\frac{3}{4}}\,(\frac{1}{\pi
\alpha^2_{\lambda}})^{\frac{3}{4}}\Big],\\
\end{eqnarray*}

\begin{eqnarray*}\Delta_{0,1}&=&(\frac{1}{\pi
\alpha^2_{\rho}})^{\frac{3}{4}}\,(\frac{1}{\pi
\alpha^2_{\lambda}})^{\frac{3}{4}}(\frac{R^2}{\pi})^{\frac{3}{4}}(\frac{\pi^2}{\lambda_1f_1})^{\frac{3}{2}}
\nonumber\\&&\exp\Big[-(f_3-\frac{f^2_2}{4f_1})|\textbf{p}|^2\Big]\nonumber\\&&\times\Big[\frac{3i}{4\pi}(\frac{1}{\pi
\alpha^2_{\rho}})^{\frac{3}{4}}(\frac{8}{3\sqrt{\pi}})^{\frac{1}{2}}(\frac{1}{
\alpha^2_{\lambda}})^{\frac{5}{4}}\Big],\\
\end{eqnarray*}

\begin{eqnarray*}\Delta_{1,0}&=&(\frac{1}{\pi
\alpha^2_{\rho}})^{\frac{3}{4}}\,(\frac{1}{\pi
\alpha^2_{\lambda}})^{\frac{3}{4}}(\frac{R^2}{\pi})^{\frac{3}{4}}(\frac{\pi^2}{\lambda_1f_1})^{\frac{3}{2}}
\nonumber\\&&\exp\Big[-(f_3-\frac{f^2_2}{4f_1})|\textbf{p}|^2\Big]\nonumber\\&&\times\Big[\frac{3i}{4\pi}(\frac{8}{3\sqrt{\pi}})^{\frac{1}{2}}(\frac{1}{
\alpha^2_{\rho}})^{\frac{5}{4}}(\frac{1}{\pi
\alpha^2_{\lambda}})^{\frac{3}{4}}\Big],\\
\end{eqnarray*}

\begin{eqnarray*}\Delta_{0,2}&=&(\frac{1}{\pi
\alpha^2_{\rho}})^{\frac{3}{4}}\,(\frac{1}{\pi
\alpha^2_{\lambda}})^{\frac{3}{4}}(\frac{R^2}{\pi})^{\frac{3}{4}}(\frac{\pi^2}{\lambda_1f_1})^{\frac{3}{2}}
\nonumber\\&&\exp\Big[-(f_3-\frac{f^2_2}{4f_1})|\textbf{p}|^2\Big]\nonumber\\&&\times\Big[\frac{\sqrt{15}}{8\pi}(\frac{1}{\pi
\alpha^2_{\rho}})^{\frac{3}{4}}(\frac{16}{15\sqrt{\pi}})^{\frac{1}{2}}(\frac{1}{
\alpha^2_{\lambda}})^{\frac{7}{4}}\Big],\\
\end{eqnarray*}

\begin{eqnarray*}\Delta_{2,0}&=&(\frac{1}{\pi
\alpha^2_{\rho}})^{\frac{3}{4}}\,(\frac{1}{\pi
\alpha^2_{\lambda}})^{\frac{3}{4}}(\frac{R^2}{\pi})^{\frac{3}{4}}(\frac{\pi^2}{\lambda_1f_1})^{\frac{3}{2}}
\nonumber\\&&\exp\Big[-(f_3-\frac{f^2_2}{4f_1})|\textbf{p}|^2\Big]\nonumber\\&&\times\Big[\frac{\sqrt{15}}{8\pi}(\frac{16}{15\sqrt{\pi}})^{\frac{1}{2}}(\frac{1}{
\alpha^2_{\rho}})^{\frac{7}{4}}(\frac{1}{\pi
\alpha^2_{\lambda}})^{\frac{3}{4}}\Big],\\
\end{eqnarray*}

\begin{eqnarray*}\Delta_{1,1}&=&(\frac{1}{\pi
\alpha^2_{\rho}})^{\frac{3}{4}}\,(\frac{1}{\pi
\alpha^2_{\lambda}})^{\frac{3}{4}}(\frac{R^2}{\pi})^{\frac{3}{4}}(\frac{\pi^2}{\lambda_1f_1})^{\frac{3}{2}}
\nonumber\\&&\exp\Big[-(f_3-\frac{f^2_2}{4f_1})|\textbf{p}|^2\Big]\nonumber\\&&\times\Big[-(\frac{3}{4\pi})^{\frac{3}{2}}\frac{8}{3\sqrt{\pi}}(\frac{1}{
\alpha^2_{\lambda}\alpha^2_{\rho}})^{\frac{5}{4}}\Big].
\end{eqnarray*}
In the above expressions, $|\textbf{p}|$ reads as
$$|\textbf{p}|=\frac{\sqrt{(m^2_A-(m_B+m_C)^2)(m^2_A-(m_B-m_C)^2)}}{2\,m_A}.$$

\vfill

\section*{Acknowledgments}
C.C. thanks W.J. Fu for the help in the numerical calculation and
Y.R. Liu and B. Zhang for useful discussions. This project was
supported by the National Natural Science Foundation of China
under Grants 10421503 and 10625521, Chinese Ministry of Education
and the China Postdoctoral Science foundation (No. 20060400376).


\begin{widetext}

\begin{center}
\begin{table}     
\caption{The decay widths of $\Lambda^{+}_{c}(2880)$  with different
D-wave assignments. All results are in units of MeV.\label{2880}}
\vskip 0.3cm
\begin{tabular}{c||ccccccccccc} \hline
 Assignment & \,\,$\Sigma^{0,+,++}_{c} \pi^{+,0,-}$ \,\,\,\,\,&
$\Sigma^{\star0,+,++}_{c} \pi^{+,0,-}$\,\,\,\,\,
&$\frac{\Gamma(\Sigma^{\star}_{c} \pi^{\pm})}{\Gamma(\Sigma_{c}
\pi^{\pm})}$\,\,\,\,\,\,\,\,\,
&$D^{0}p$\,\,\,\,\,\,\,\,&Remark\\
\hline \hline

 $\Lambda_{c2}(\frac{3}{2}^{+})$&$7.8$&$0.9$&$ 0.11$&$0.0$&$\times$ \\
 $\Lambda_{c2}(\frac{5}{2}^{+})$&$0.06$&$5.34$&$89 $&$0.0$&$\times$\\
 \hline
 $\hat{\Lambda}_{c2}(\frac{3}{2}^{+})$&$78.3$&$59.1$&$0.75 $&$0.0$&$\times $\\
 $\hat{\Lambda}_{c2}(\frac{5}{2}^{+})$&$78.3$&$59.1$&$0.75 $&$0.0 $&$\times $\\
 \hline

 $\check{\Lambda}^{0}_{c1}(\frac{1}{2}^{+})$&$0.9$&$2.3$&$2.6 $&$ 2.3$&$\times$\\
 $\check{\Lambda}^{0}_{c1}(\frac{3}{2}^{+})$&$ 0.22$&$6.0$&$ 27$&$ 2.3$&$\times $\\

 $\check{\Lambda}^{1}_{c0}(\frac{1}{2}^{+})$&$132$&$ 144$&$ 1.1$&$ 0.0$&$\times$\\
 $\check{\Lambda}^{1}_{c1}(\frac{1}{2}^{+})$&$ 66.3$&$ 18.0$&$ 0.27$&$150 $&$\times $\\
 $\check{\Lambda}^{1}_{c1}(\frac{3}{2}^{+})$&$16.5$&$ 45.0$&$ 2.7$&$ 150$&$\times$\\
 $\check{\Lambda}^{1}_{c2}(\frac{3}{2}^{+})$&$ 82.8$&$ 9.0$&$ 0.10$&$0.0 $&$\times $\\
 $\check{\Lambda}^{1}_{c2}(\frac{5}{2}^{+})$&$ 0.0$&$ 54.1$&$-$&$0.0 $&$\times$\\

 $\check{\Lambda}^{2}_{c1}(\frac{1}{2}^{+})$&$ 25.7$&$8.1$&$0.32 $&$64 $&$\times $\\
 $\check{\Lambda}^{2}_{c1}(\frac{3}{2}^{+})$&$ 6.5$&$20.4$&$ 3.1$&$ 64$&$\times$\\
 $\check{\Lambda}^{2}_{c2}(\frac{3}{2}^{+})$&$ 57.9$&$14.2$&$ 0.24$&$0.0$&$\times $\\
 $\check{\Lambda}^{2}_{c2}(\frac{5}{2}^{+})$&$ 9.4$&$ 47.1$&$ 5.0$&$0.0$&$\times $\\
 $\check{\Lambda}^{2}_{c3}(\frac{5}{2}^{+})$&$ 10.8$&$5.5$&$ 0.51$&$12$\\
 $\check{\Lambda}^{2}_{c3}(\frac{7}{2}^{+})$&$ 6.1$&$ 7.4$&$ 1.2$&$12$&$\times $\\
 \hline
 \hline
\end{tabular}
\end{table}
\begin{table}      
\caption{The decay widths of $\Lambda^{+}_{c}(2940)$  with different
D-wave assignments. Here all results are in units of MeV.
\label{2940}}\vskip 0.3cm
\begin{tabular}{c||ccccccccccc}
\hline
 Assignment & \,\,$\Sigma^{0,+,++}_{c} \pi^{+,0,-}$ \,\,\,\,\,&
\;\;\;$\Sigma^{\star0,+,++}_{c} \pi^{+,0,-}$\,\,\,\,\,
\,\,&$\frac{\Gamma(\Sigma^{\star}_{c} \pi^{\pm})}{\Gamma(\Sigma_{c}
\pi^{\pm})}$\,\,\,\,\,\,&$D^{0}p$\,\,\,\,\,\,&Remark \\\hline \hline

           $\Lambda_{c2}(\frac{3}{2}^{+})$&$11.7$&$9.1$&$0.77$&$0.0 $&$\times$\\
             $\Lambda_{c2}(\frac{5}{2}^{+})$&$0.2$&$9.1$&$46$&$ 0.0$&$\times$\\
\hline

       $\hat{\Lambda}_{c2}(\frac{3}{2}^{+})$&$170$  &  $150$&$0.88$&$0.0 $&$\times$\\
       $\hat{\Lambda}_{c2}(\frac{5}{2}^{+})$&$170$  &  $150$&$0.88$&$0.0 $&$\times$\\
\hline

 $\check{\Lambda}^{0}_{c1}(\frac{1}{2}^{+})$&$2.2$  &  $0.5$&$0.23$&$11 $&$$\\
 $\check{\Lambda}^{0}_{c1}(\frac{3}{2}^{+})$&$0.6 $  &  $1.4$&$2.3$&$11 $&$$\\

 $\check{\Lambda}^{1}_{c0}(\frac{1}{2}^{+})$&$212$  &  $ 259$&$1.2$&$0.0 $&$\times$\\
 $\check{\Lambda}^{1}_{c1}(\frac{1}{2}^{+})$&$106$  &  $ 32.4$&$0.31$&$340 $&$\times$\\
 $\check{\Lambda}^{1}_{c1}(\frac{3}{2}^{+})$&$26.5$&  $81.0$&$3.1$&$340 $&$\times$\\
 $\check{\Lambda}^{1}_{c2}(\frac{3}{2}^{+})$&$142$  & $16.2$&$0.11$&$0.0 $&$\times$\\
 $\check{\Lambda}^{1}_{c2}(\frac{5}{2}^{+})$&$ 0.0   $  & $97.0$&$-$&$0.0 $&$\times$\\

 $\check{\Lambda}^{2}_{c1}(\frac{1}{2}^{+})$&$34.5$ &  $12.6$&$0.37$&$95$&$\times$\\
 $\check{\Lambda}^{2}_{c1}(\frac{3}{2}^{+})$&$8.6 $  &  $ 31.7$&$3.7$&$95$&$\times$\\
 $\check{\Lambda}^{2}_{c2}(\frac{3}{2}^{+})$&$77.7$  &  $ 27.7$&$0.36$&$0.0$&$\times$\\
 $\check{\Lambda}^{2}_{c2}(\frac{5}{2}^{+})$&$19.5$  &  $75.6$&$3.9$&$0.0$&$\times$\\
 $\check{\Lambda}^{2}_{c3}(\frac{5}{2}^{+})$&$22.2$  &  $12.9$&$0.58$&$49$&$\times$\\
 $\check{\Lambda}^{2}_{c3}(\frac{7}{2}^{+})$&$12.4$  &   $17.5$&$1.4$&$49$&$\times$\\
 \hline
 \hline
\end{tabular}
\end{table}
\end{center}


\begin{center}
\begin{table}[htb]      
\caption{The decay widths of $\Xi^{+}_{c}(2980)$  with different
D-wave assignments. Here all results are in units of MeV.
\label{2980}}\vskip 0.3cm
\begin{tabular}{c||ccccccccccc}
\hline
 Assignment & \,\,$\Xi^{0}_{c}
\pi^{+}$&\,\,$\Xi'^{0}_{c} \pi^{+}$ & \,\,$\Xi^{\star0}_{c}
\pi^{+}$&\,\,$\Sigma^{++}_{c}k^{-}$&\,\,\,$\Lambda^{+}_{c}\bar{k}^{0}$&Remark
\\\hline\hline

 $\Xi_{c2}(\frac{3}{2}^{+})$             &$0.0$&  $1.1$  &$0.11$  &$0.37$  &$0.0$&$\times$\\
 $\Xi_{c2}(\frac{5}{2}^{+})$             &$0.0$&  $0.12\times10^{-2}$  &$0.67$  &$0.11\times10^{-3}$ &$0.0$&$\times$\\
 \\
 $\Xi'_{c1}(\frac{1}{2}^{+})$            &$4.4$&  $0.72$  &$0.18$  &$0.25$  &$5.3$&$$\\
 $\Xi'_{c1}(\frac{3}{2}^{+})$            &$4.4$&  $0.18$  &$0.46$  &$0.062$  &$5.3$&$$\\

 $\Xi'_{c2}(\frac{3}{2}^{+})$            &$0.0$&  $0.16$  &$0.17$  &$0.56$  &$0.0$&$\times$\\
 $\Xi'_{c2}(\frac{5}{2}^{+})$            &$0.0$&  $0.47\times10^{-2}$  &$1.0$  &$0.71\times10^{-4}$  &$0.0$&$\times$\\

 $\Xi'_{c3}(\frac{5}{2}^{+})$&$0.054$&$0.53\times10^{-2}$&$0.14\times10^{-2}$&
 $0.82\times10^{-4}$  &$0.053$&$\times$\\
 $\Xi'_{c3}(\frac{7}{2}^{+})$&$0.054$&  $0.30\times10^{-2}$
 &$0.19\times10^{-2}$  &$0.46\times10^{-4}$  &$0.053$&$\times$\\
  \hline
 $\hat{\Xi}_{c2}(\frac{3}{2}^{+})$       &$0.0$&  $9.5$  &$6.1$  &$0.61$  &$0.0$&$$\\
 $\hat{\Xi}_{c2}(\frac{5}{2}^{+})$       &$0.0$&  $9.5$  &$6.1$  &$0.61$  &$0.0$&$$\\
 \\
 $\hat{\Xi'}_{c1}(\frac{1}{2}^{+})$      &$74$&  $6.3$  &$1.0$  &$0.40$  &$78$&$\times$\\
 $\hat{\Xi'}_{c1}(\frac{3}{2}^{+})$      &$74$&  $1.6$  &$2.5$  &$0.10$  &$78$&$\times$\\

 $\hat{\Xi'}_{c2}(\frac{3}{2}^{+})$      &$0.0$&  $14$  &$4.5$  &$0.91$  &$0.0$&$$\\
 $\hat{\Xi'}_{c2}(\frac{5}{2}^{+})$      &$0.0$&  $6.3$  &$7.1$  &$0.40$  &$0.0$&$$\\

 $\hat{\Xi'}_{c3}(\frac{5}{2}^{+})$      &$48$&  $7.2$  &$2.9$  &$0.46$  &$50$&$\times$\\
 $\hat{\Xi'}_{c3}(\frac{7}{2}^{+})$      &$48$&  $4.1$  &$3.9$  &$0.26$  &$50$&$\times$\\
 \hline
 $\check{\Xi'}^{0}_{c0}(\frac{1}{2}^{+}) $&$0.0$& $0.30$  &$1.4$  &$1.3$  &$0.0$&$\times$\\
 $\check{\Xi}^{0}_{c1}(\frac{1}{2}^{+})$ &$1.0$& $0.40$  &$0.46$  &$1.7$  &$0.46$&$\times$\\
 $\check{\Xi}^{0}_{c1}(\frac{3}{2}^{+})$ &$1.0$& $0.10$  &$1.2$  &$0.43$  &$0.46$&$\times$\\
 \\
 $\check{\Xi'}^{1}_{c1}(\frac{1}{2}^{+})$&$0.0$&  $18$  &$4.4$  &$5.5$  &$0.0$&$$\\
 $\check{\Xi'}^{1}_{c1}(\frac{3}{2}^{+})$&$0.0$&  $4.5$  &$11$  &$1.4$  &$0.0$&$$\\
 $\check{\Xi}^{1}_{c0}(\frac{1}{2}^{+})$ &$0.0$&  $18$  &$18$  &$5.5$  &$0.0$&$$\\
 $\check{\Xi}^{1}_{c1}(\frac{1}{2}^{+})$ &$62$& $9.1$  &$2.2$  &$2.8$  &$72$&$\times$\\
 $\check{\Xi}^{1}_{c1}(\frac{3}{2}^{+})$ &$62$& $2.3$  &$5.5$  &$0.69$  &$72$&$\times$\\
 $\check{\Xi}^{1}_{c2}(\frac{3}{2}^{+})$ &$0.0$&  $11$  &$1.1$  &$0.34$  &$0.0$&$\times$\\
 $\check{\Xi}^{1}_{c2}(\frac{5}{2}^{+})$ &$0.0$&$0.0$  &$6.6$  &$0.0$  &$0.0$&$\times$\\
 \\
 $\check{\Xi'}^{2}_{c2}(\frac{3}{2}^{+})$&$0.0$&  $5.6$  &$1.8$  &$2.4$  &$0.0$&$$\\
 $\check{\Xi'}^{2}_{c2}(\frac{5}{2}^{+})$&$0.0$&  $1.7$  &$4.32$  &$0.24$  &$0.0$&$$\\
 $\check{\Xi}^{2}_{c1}(\frac{1}{2}^{+})$ &$19$&  $3.7$  &$1.1$  &$1.6$  &$23$&$$\\
 $\check{\Xi}^{2}_{c1}(\frac{3}{2}^{+})$ &$19$&  $0.93$  &$2.6$  &$0.40$  &$23$&$$\\
 $\check{\Xi}^{2}_{c2}(\frac{3}{2}^{+})$ &$0.0$&  $8.4$  &$1.7$  &$0.36$  &$0.0$&$$\\
 $\check{\Xi}^{2}_{c2}(\frac{5}{2}^{+})$ &$0.0$&  $1.2$  &$6.0$  &$0.16$  &$0.0$&$$\\
 $\check{\Xi}^{2}_{c3}(\frac{5}{2}^{+})$ &$8.1$&  $1.3$  &$60$  &$0.19$  &$8.7$&$\times$\\
 $\check{\Xi}^{2}_{c3}(\frac{7}{2}^{+})$ &$8.1$&  $0.75$  &$0.81$  &$0.10$  &$8.7$&$$\\
 \hline\hline
\end{tabular}
\end{table}

\begin{table}      
\caption{The decay widths of $\Xi^{+}_{c}(3077)$  with different
D-wave assignments. Here all results are in units of MeV.
\label{3077} }\vskip 0.3cm
\begin{tabular}{c||ccccccccccc}
\hline
 Assignment & \,\,$\Xi^{0}_{c}
\pi^{+}$&\,\,$\Xi'^{0}_{c} \pi^{+}$ & \,\,$\Xi^{\star0}_{c}
\pi^{+}$&\,\,$\Sigma^{++}_{c}k^{-}$&\,\,$\Sigma^{++}_{c}k^{-}$&\,$\Lambda^{+}_{c}\bar{k}^{0}$
&\;\;\;$D^{+}\Lambda$&\;\;\; Remark\\\hline\hline

 $\Xi_{c2}(\frac{3}{2}^{+})$             &$0.0$&  $2.1$  &$0.30$  &$0.73$  &$0.054$  &$0.0$&$0.0$&$$\\
 $\Xi_{c2}(\frac{5}{2}^{+})$             &$0.0$&  $0.037$  &$1.7$  &$0.42\times10^{-2}$  &$0.32$  &$0.0$&$0.0$&$$\\
 \\
 $\Xi'_{c1}(\frac{1}{2}^{+})$            &$7.0$&  $1.4$  &$0.46$  &$0.49$  &$0.089$  &$4.4$&$3.2$&\\
 $\Xi'_{c1}(\frac{3}{2}^{+})$            &$7.0$&  $0.36$  &$1.1$  &$0.12$  &$0.22$  &$4.4$&$3.2$&\\

 $\Xi'_{c2}(\frac{3}{2}^{+})$            &$0.0$&  $3.2$  &$0.43$  &1.1$$  &$0.081$  &$0.0$&$0.0$&$$\\
 $\Xi'_{c2}(\frac{5}{2}^{+})$            &$0.0$&  $0.025\times10^{-2}$  &$2.5$  &$0.28\times10^{-2}$  &$0.48$  &$0.0$&$0.0$&$\times$\\

 $\Xi'_{c3}(\frac{5}{2}^{+})$&$0.19$&$0.029$&$0.012$
  &$0.32\times10^{-2}$&$0.32\times10^{-3}$  &$0.12$&$0.026$&$\times$\\
 $\Xi'_{c3}(\frac{7}{2}^{+})$&$0.19$&$0.016\times10^{-3}$
 &$0.016$&$0.18\times10^{-2}$&$0.44\times10^{-3}$  &$0.12$&$0.026$&$\times$\\
  \hline

 $\hat{\Xi}_{c2}(\frac{3}{2}^{+})$       &$0.0$&  $34$  &$29$  &$6.0$  &$2.0$  &$0.0$&$0.0$&$\times$\\
 $\hat{\Xi}_{c2}(\frac{5}{2}^{+})$       &$0.0$&  $34$  &$29$  &$6.0$  &$2.0$  &$0.0$&$0.0$&$\times$\\
 \\
 $\hat{\Xi'}_{c1}(\frac{1}{2}^{+})$      &$201$&  $23$  &$4.8$  &$4.0$  &$0.33$  &$130$&$38$&$\times$\\
 $\hat{\Xi'}_{c1}(\frac{3}{2}^{+})$      &$201$&  $5.7$  &$12$  &$1.0$  &$0.83$  &$130$&$38$&$\times$\\

 $\hat{\Xi'}_{c2}(\frac{3}{2}^{+})$      &$0.0$&  $51$  &$22$  &$8.9$  &$1.5$  &$0.0$&$0.0$&$\times$\\
 $\hat{\Xi'}_{c2}(\frac{5}{2}^{+})$      &$0.0$&  $23$  &$34$  &$4.0$  &$2.3$  &$0.0$&$0.0$&$\times$\\

 $\hat{\Xi'}_{c3}(\frac{5}{2}^{+})$      &$129$&  $26$  &$14$  &$4.5$  &$0.94$ &$84$&$25$&$\times$\\
 $\hat{\Xi'}_{c3}(\frac{7}{2}^{+})$      &$129$&  $15$  &$19$  &$2.6$  &$0.13$ &$84$&$25$&$\times$\\
 \hline

 $\check{\Xi'}^{0}_{c0}(\frac{1}{2}^{+}) $&$0.0$&  $0.69$  &$0.13$  &$0.29$ &$1.2$ &$0.0$&$0.0$&$$\\
 $\check{\Xi}^{0}_{c1}(\frac{1}{2}^{+})$ &$15$&  $0.92$  &$0.044$  &$0.39$  &$0.38$ &$11$&$0.64\times10^{-3}$&$\times$\\
 $\check{\Xi}^{0}_{c1}(\frac{3}{2}^{+})$ &$15$&  $0.23$  &$0.11$  &$0.096$&$0.96$ &$11$&$0.64\times10^{-3}$&$\times$\\
 \\
 $\check{\Xi'}^{1}_{c1}(\frac{1}{2}^{+})$&$0.0$&  $39$  &$12$  &$12$  &$0.21$ &$0.0$&$0.0$&$\times$\\
 $\check{\Xi'}^{1}_{c1}(\frac{3}{2}^{+})$&$0.0$&  $9.9$  &$30$  &$3.0$  &$5.2$ &$0.0$&$0.0$&$\times$\\
 $\check{\Xi}^{1}_{c0}(\frac{1}{2}^{+})$ &$0.0$&  $39$  &$47$  &$12$  &$8.3$ &$0.0$&$0.0$&$\times$\\
 $\check{\Xi}^{1}_{c1}(\frac{1}{2}^{+})$ &$110$&  $20$  &$5.9$  &$6.1$  &$1.0$ &$69$&$42$&$\times$\\
 $\check{\Xi}^{1}_{c1}(\frac{3}{2}^{+})$ &$110$&  $5.0$  &$15$  &$1.5$  &$2.6$ &$69$&$42$&$\times$\\
 $\check{\Xi}^{1}_{c2}(\frac{3}{2}^{+})$ &$0.0$&  $25$  &$3.0$  &$7.6$  &$0.52$ &$0.0$&$0.0$&$\times$\\
 $\check{\Xi}^{1}_{c2}(\frac{5}{2}^{+})$ &$0.0$&  $0.0$  &$18$  &$0.0$  &$3.1$ &$0.0$&$0.0$&$\times$\\
 \\
 $\check{\Xi'}^{2}_{c2}(\frac{3}{2}^{+})$&$0.0$&  $9.2$  &$6.0$  &$3.9$  &$0.75$ &$0.0$&$0.0$&$$\\
 $\check{\Xi'}^{2}_{c2}(\frac{5}{2}^{+})$&$0.0$&  $5.8$  &$10$  &$1.1$  &$2.1$ &$0.0$&$0.0$&$$\\
 $\check{\Xi}^{2}_{c1}(\frac{1}{2}^{+})$ &$22$&  $6.1$  &$2.3$  &$2.6$  &$0.54$ &$14$&$15$&$\times$\\
 $\check{\Xi}^{2}_{c1}(\frac{3}{2}^{+})$ &$22$&  $1.5$  &$5.6$  &$0.64$  &$1.3$ &$14$&$15$&$\times$\\
 $\check{\Xi}^{2}_{c2}(\frac{3}{2}^{+})$ &$0.0$&  $14$  &$5.2$  &$5.8$  &$0.77$ &$0.0$&$0.0$&$\times$\\
 $\check{\Xi}^{2}_{c2}(\frac{5}{2}^{+})$ &$0.0$&  $3.9$  &$14$  &$0.74$  &$3.0$ &$0.0$&$0.0$&$\times$\\
 $\check{\Xi}^{2}_{c3}(\frac{5}{2}^{+})$ &$21$&  $4.4$  &$2.5$  &$0.85$  &$0.23$ &$14$&$4.3$&$\times$\\
 $\check{\Xi}^{2}_{c3}(\frac{7}{2}^{+})$ &$21$&  $2.5$  &$3.4$  &$0.48$  &$0.31$ &$14$&$4.3$&$\times$\\
 \hline
 \hline
\end{tabular}
\end{table}
\end{center}

\begin{figure}[htb]
\begin{center}
\begin{tabular}{ccc}
\scalebox{0.55}{\includegraphics{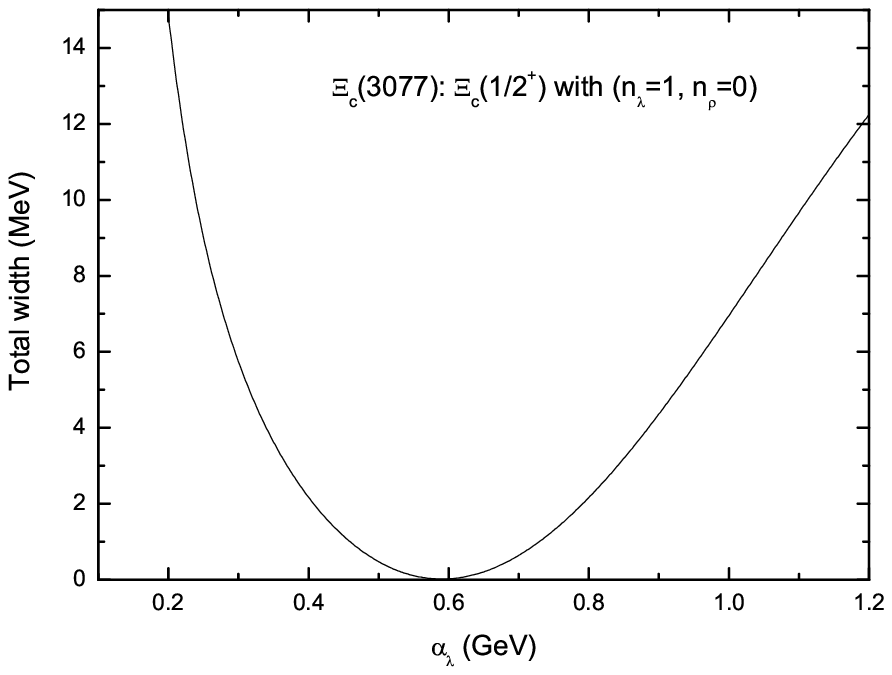}}&\scalebox{0.55}{\includegraphics{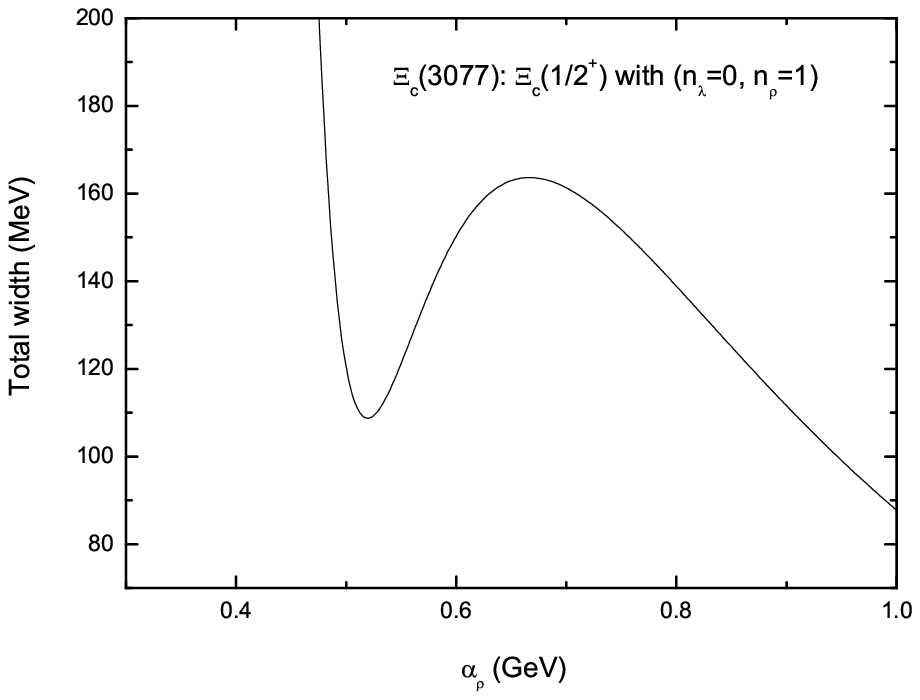}}&\scalebox{0.55}{\includegraphics{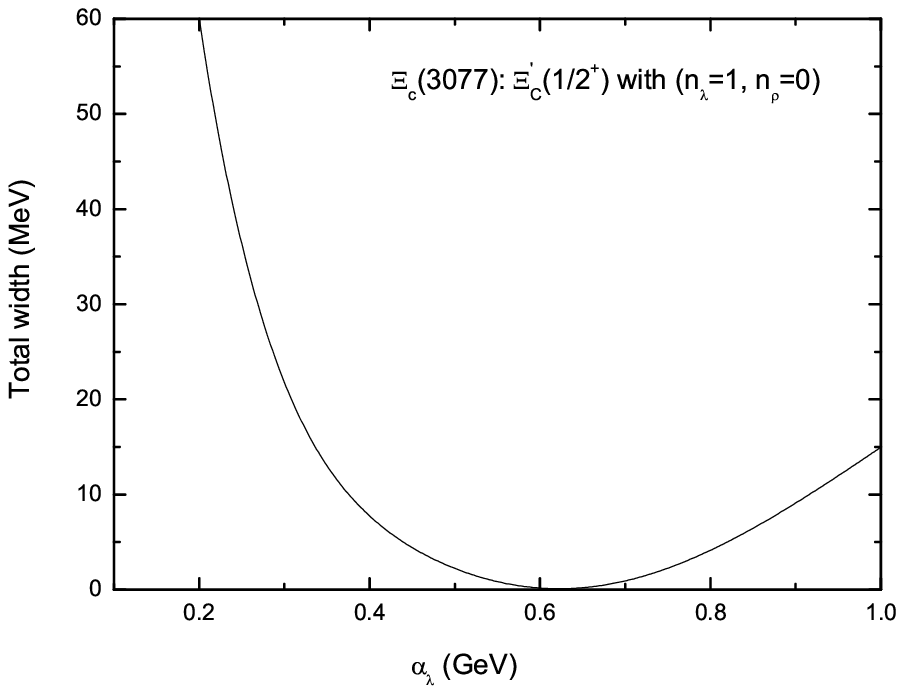}}\\
(a)&(b)&(c)\\
\scalebox{0.55}{\includegraphics{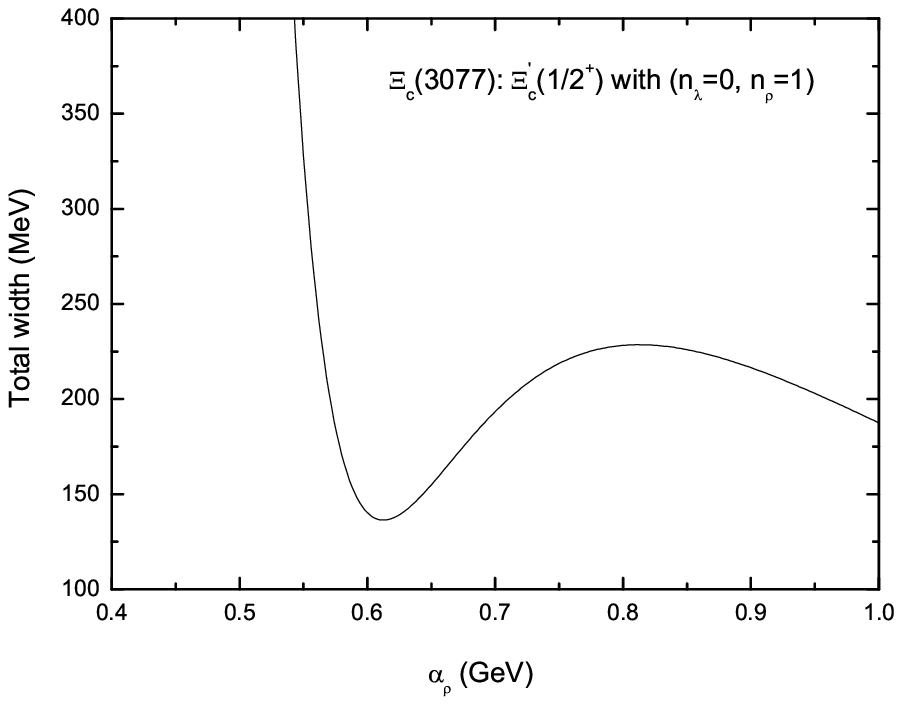}}&\scalebox{0.55}{\includegraphics{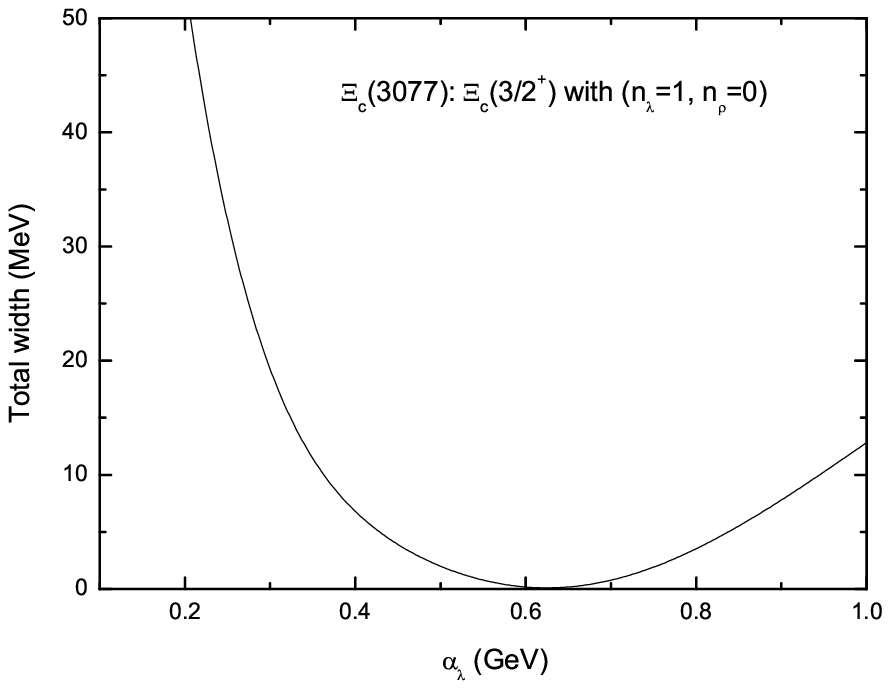}}&\scalebox{0.55}{\includegraphics{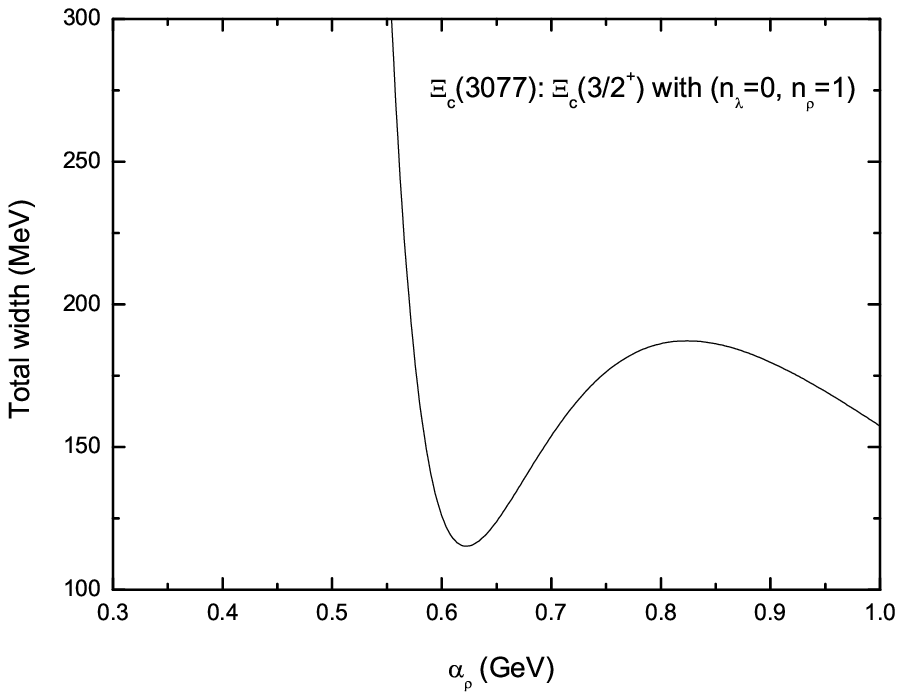}}\\
(d)&(e)&(f)\\
\end{tabular}
\end{center}
\caption{The dependence of the total decay width of
$\Xi_{c}(3077)$ on the parameter $\alpha_{\lambda}$ or
$\alpha_{\rho}$ if $\Xi_{c}(3077)$ is a radial excitation. In
these figures, we fix $\alpha_{\rho}=0.6$ GeV for the case with
$n_{\lambda}=1$, and $\alpha_{\lambda}=0.6$ GeV for $n_{\rho}=1$.
The situation of $\Xi_{c}(2980)$ as a radial excitation is very
similar. \label{S-wave-radical-1}}
\end{figure}

\end{widetext}


\begin{thebibliography}{99}

\bibitem{babar-2880}BABAR Collaboration, B. Aubert et al., Phys. Rev. Lett. {\bf 98}, 012001
(2007).

\bibitem{belle-2880}BELLE Collaboration, K. Abe et al., arXiv:
hep-ex/0608043.

\bibitem{babar-2980-3077}BABAR Collaboration, B. Aubert et al.,
arXiv: hep-ex/0607042.

\bibitem{belle-2980-3077}BELLE Collaboration, R. Chistov et al.,
Phys. Rev. Lett. {\bf 97}, 162001 (2006).

\bibitem{babar-omega}BABAR Collaboration, B. Aubert et al., arXiv:
hep-ex/0608055.

\bibitem{rosner}J.L. Rosner, arXiv: hep-ph/0612332; arXiv: hep-ph/0609195; arXiv:
hep-ph/0606166.
\bibitem{xiang}X.G. He, Xue-Qian Li, Xiang Liu and X.Q. Zeng,
arXiv: hep-ph/0606015.

\bibitem{cheng}H.Y. Cheng and C.K. Chua, Phys. Rev. {\bf D 75},
014006 (2007).

\bibitem{valcarce}H. Garcilazo, J. Vijande and A. Valcarce, arXiv:
hep-ph/0703257.

\bibitem{cleo-2880}CLEO Collaboration, M. Artuso et al., Phys. Rev.
Lett. {\bf 86}, 4479 (2001).

\bibitem{charmed baryons}S. Tawfiq, P.J. O'Donnell, and J.G. K\"{o}rner, Phys. Rev. {\bf D 58}, 054010
(1998); M.A. Ivanov, J.G. K\"{o}rner, V.E. Lyubovitskij, and A.G.
Rusetsky, Phys. Rev. {\bf D 60}, 094002 (1999); M.Q. Huang, Y.B.
Dai, and C.S. Huang, Phys. Rev. {\bf D 52}, 3986 (1995); ibid. {\bf
D 55}, 7317(E) (1997); S.L. Zhu, Phys. Rev. {\bf D 61}, 114019
(2000).
\bibitem{review}D. Pirjol and T.M. Yan, Phys. Rev. {\bf D 56}, 5483
(1997).
\bibitem{PDG}W.M. Yao et al., Particle Data Group, J. Phys. G {\bf 33}, 1
(2006).

\bibitem{CDF}CDF Collaboration, I.V.Gorelov, arXiv: hep-ex/0701056.
\bibitem{CDF-1}I.V. Gorelov, arXiv: hep-ex/0701056.

\bibitem{theory}E. Jenkins, Phys. Rev. {\bf D 54}, 4515 (1996);
ibid. {\bf 55}, 10 (1997); M. Karlinear and H.J. Lipkin, arXiv:
hep-ph/0307243; M. Karlinear and H.J. Lipkin, Phys. Lett. {\bf B
575}, 249 (2003).

\bibitem{rosner-1}J.L. Rosner, Phys. Rev. {\bf D 75}, 013009 (2007).
\bibitem{lipkin}M. Karliner and H.J. Lipkin, arXiv: hep-ph/0611306.
\bibitem{hwang}C.W. Hwang, arXiv: hep-ph/0611221.



\bibitem{Micu} L. Micu, Nucl. Phys. {\bf B10}, 521 (1969).


\bibitem{yaouanc}A. Le Yaouanc, L. Oliver, O. P\`{e}ne and J. Raynal,
Phys. Rev. {\bf D8}, 2223 (1973); {\bf D9}, 1415 (1974); {\bf D11},
1272 (1975); Phys. lett. {\bf B71}, 57 (1977); {\bf B71}, 397
(1977).

\bibitem{yaouanc-1}A. Le Yaouanc, L. Oliver, O. P\`{e}ne and J. Raynal,
Phys. Lett. {\bf B72}, 57 (1977).

\bibitem{yaouanc-book} A. Le Yaouanc, L. Oliver, O. P\`{e}ne and J. Raynal,
{\it Hadron Transitions in the Quark Model}, Gordon and Breach
Science Publishers, New York, 1987.

\bibitem{qpc-1} H.G. Blundell and S. Godfrey, Phys. Rev. {\bf  D53}, 3700
(1996).
\bibitem{qpc-2}P.R. Page, Nucl. Phys. {\bf B446}, 189 (1995); S. Capstick and N.
Isgur, Phys. Rev. {\bf D34}, 2809 (1986).

\bibitem{qpc-90}S. Capstick and W. Roberts, Phys. Rev.  {\bf D49}, 4570 (1994).
\bibitem{ackleh}E.S. Ackleh, T. Barnes and E.S. Swanson, Phys. Rev.  {\bf D54}, 6811 (1996).
\bibitem{Zou}H.Q. Zhou, R.G. Ping and B.S. Zou, Phys. Lett. {\bf B611}, 123
(2005).
\bibitem{liu}X.H. Guo, H.W. Ke, X.Q. Li, X. Liu and S.M. Zhao,
arXiv: hep-ph/0510146.
\bibitem{lujie}J. Lu, W.Z. Deng, X.L. Chen and S.L. Zhu,  Phys. Rev. {\bf D 73} 054012,
(2006); B. Zhang, X. Liu and S.L. Zhu, DOI:
10.1140/epjc/s10052-007-0221-y, arXiv: hep-ph/0609013.
\bibitem{baryon-decay}S. Capstick and W. Roberts, Phys. Rev. {\bf D
47}, 1994 (1993).

\bibitem{mockmeson}
C. Hayne and N. Isgur, Phys. Rev. D {\bf 25}, 1944 (1982).

\bibitem{potential}S. Capstick and N. Isgur, Phys. Rev. {\bf D 34},
2809 (1986).

\bibitem{Godfrey}H. G. Blundell, S. Godfrey, Phys. Rev. {\bf D 53}, 3700 (1996).




\bibitem{parameter-2}F.E. Close and E.S. Swanson, Phys. Rev. {\bf D 72}, 094004
(2005).



\bibitem{E.Jenkins}E. Jenkins, Phys. Rev. {\bf D 54},
4515 (1996).
\bibitem{belle-2800}Belle Collaboration, R. Mizuk, Phys. Rev. Lett. {\bf 94},
122002 (2005).


\end{thebibliography}
\end{document}